# Review of Heavy-Ion Inertial Fusion Physics

S. Kawata[1, 2, *], T. Karino[1], and A. I. Ogoyski[3]

[1]*Graduate School of Engineering, Utsunomiya University, Yohtoh 7-1-2, Utsunomiya 321-8585, Japan*
[2]*CORE (Center for Optical Research and Education), Utsunomiya University, Yohtoh 7-1-2, Utsunomiya 321-8585, Japan*
[3]*Department of Physics, Technical University of Varna, Ulitska, Studentska 1, Varna, Bulgaria*
*Corresponding author: kwt@cc.utsunomiya-u.ac.jp

In this review paper on heavy ion inertial fusion (HIF), the state-of-the-art scientific results are presented and discussed on the HIF physics, including physics of the heavy ion beam (HIB) transport in a fusion reactor, the HIBs-ion illumination on a direct-drive fuel target, the fuel target physics, the uniformity of the HIF target implosion, the smoothing mechanisms of the target implosion non-uniformity and the robust target implosion. The HIB has remarkable preferable features to release the fusion energy in inertial fusion: in particle accelerators HIBs are generated with a high driver efficiency of ~ 30-40%, and the HIB ions deposit their energy inside of materials. Therefore, a requirement for the fusion target energy gain is relatively low, that would be ~50-70 to operate a HIF fusion reactor with the standard energy output of 1GW of electricity. The HIF reactor operation frequency would be ~10~15 Hz or so. Several-MJ HIBs illuminate a fusion fuel target, and the fuel target is imploded to about a thousand times of the solid density. Then the DT fuel is ignited and burned. The HIB ion deposition range is defined by the HIB ions stopping length, which would be ~1 mm or so depending on the material. Therefore, a relatively large density-scale length appears in the fuel target material. One of the critical issues in inertial fusion would be a spherically uniform target compression, which would be degraded by a non-uniform implosion. The implosion non-uniformity would be introduced by the Rayleigh-Taylor (R-T) instability, and the large density-gradient-scale length helps to reduce the R-T growth rate. On the other hand, the large scale length of the HIB ions stopping range suggests that the temperature at the energy deposition layer in a HIF target does not reach a very-high temperature: normally the about 300eV or so is realized in the energy absorption region, and that a direct-drive target would be appropriate in HIF. In addition, the HIB accelerators are operated repetitively and stably. The precise control of the HIB axis manipulation is also realized in the HIF accelerator, and the HIB wobbling motion may give another tool to smooth the HIB illumination non-uniformity. The key issues in HIF physics are also discussed and presented in the paper.

## 1 INTRODUCTION

The heavy ion beam (HIB) fusion (HIF) has been proposed many years ago in 1970's. For example, a proton-beam inertial fusion target was proposed in Ref. [1]. The HIF reactor designs were also proposed in Refs. [2-4].

HIB ions deposit their energy inside of materials, and the interaction of the HIB ions with the materials are well understood [5, 6]. The HIB ion interaction with a material is explained and defined well by the classical Coulomb collision and a plasma wave excitation in the material plasma. The HIB ions do not reflect from the target material, and deposit all the HIB ion energy inside of the material. The HIB energy deposition length is typically the order of ~mm in an HIF fuel target depending on the HIB ion energy and the material. When several MJ of the HIB energy is deposited in the material in an inertial confinement fusion (ICF) fuel target, the temperature of the energy deposition layer plasma becomes about 300 eV or so. The peak temperature or the peak plasma pressure appears near the HIB ion stopping area by the Bragg peak effect, which comes from the nature of the Coulomb collision. The total stopping range would be normally wide and the order of ~mm inside of the material. An indirect drive target was also proposed in Ref. [7]. However, due to the relatively moderate temperature inside of the material and the plasma temperature and pressure profiles, a spherical direct drive target would be appropriate in HIF.

In ICF, a driver efficiency and its repetitive operation with several Hz ~ 20 Hz or so are essentially important to constitute an ICF reactor system. HIB driver accelerators have a high driver energy efficiency of 30-40 % from the electricity to the HIB energy. In general, high-energy accelerators have been operated repetitively daily. The high driver efficiency relaxes the requirement for the fuel target gain. In HIF the target gain of 50~70 allows us to construct HIF fusion reactor systems, and 1MkW of the electricity output would be realized with the repetition rate of 10~15 Hz.

The HIB accelerator has a high controllability to define the ion energy, the HIB pulse shape, the HIB pulse length and the HIB number density or current as well as the beam axis. The HIB axis could be also controlled or oscillated with a high frequency [8-10]. The controlled wobbling motion of the HIB axis is one of remarkable preferable points in HIF, and would contribute to smooth the HIBs illumination non-uniformity on a DT fuel target and to mitigate the Rayleigh-Taylor (R-T) instability growth in the HIF fuel target implosion [11-13].

The relatively large density gradient scale length is created in the HIBs energy deposition region in an DT fuel target, and it also contribute to reduce the R-T instability growth rate especially for shorter wavelength modes [14, 15]. So in the HIF target implosion longer wavelength modes should be focused for the target implosion uniformity.

In ICF target implosion, the requirement for the implosion uniformity is very stringent, and the implosion non-uniformity must be less than a few % [16, 17]. Therefore, it is essentially important to improve the fuel target implosion uniformity. In general the target implosion non-uniformity is introduced by a driver beams' illumination non-uniformity, an imperfect target sphericity, a non-uniform target density, a target alignment error in a fusion reactor, et al. The target implosion should be robust against the implosion non-uniformities for the stable reactor operation.

In the HIBs energy deposition region in a DT fuel target a wide density valley appears, and in the density valley a part of the HIBs deposited energy is converted to the radiation and the radiation is confined in the density valley. The converted and confined radiation energy is not negligible, and it would be the order of ~100 kJ in a HIF reactor-size DT target. The confined radiation in the density valley contributes also to reduce the non-uniformity of the HIBs energy deposition.

The HIB must be transported in a fusion reactor, which would be filled by a debris gas plasma. The reactor radius would be 3~5 m or so. From the beam exit of the accelerator the HIB should be transported stably. The HIB ion is rather heavy. For example, $Pb^+$ ion beams could be a promising candidate for the HIF driver beam. Fortunately the heavy ions are transported almost ballistically with straight trajectories in a long distance. Between the HIB ions and the background electrons, beam-plasma interactions occur: the two-stream and filamentation instabilities may appear, and simple analyses confirm that the HIBs are almost safe from the instabilities' influences. However, the HIB's self charge may contribute to a slight radial expansion of the HIBs especially near the fuel target area due to the neutralized electrons' heating by the HIB radial compression during the HIB's propagation in a reactor [18]. So the HIB charge neutralization is also discussed in the paper.

The HIB uniform illumination is also studied, and the target implosion uniformity requirement requests the minimum HIB number: details HIBs energy deposition on a direct-drive DT fuel target shows that the minimum HIBs number would be the 32 beams [19]. The 3-dimensional detail HIBs illumination on a HIF DT target is computed by a computer code of OK [20]. The HIBs illumination non-uniformity is also studied in detail. One of the study results shows that a target misalignment of ~100μm is tolerable in fusion reactor to release the HIF energy stably.

In the paper first the issues in HIF are summarized, and the key requirements are then discussed. The HIB transport is examined with discussions on the HIB neutralization and the relating instabilities. The HIB illumination uniformity and the HIF target physics are then discussed in detail. Presented are the HIB energy deposition, the HIB illumination non-uniformity, the target implosion dynamics, the implosion uniformity, the HIB ion interaction with the target and the robust target implosion. The chamber gas dynamics is also touched briefly. In addition, the HIB projects are also introduced. The summaries are presented finally.

## 2 ISSUES IN HEAVY ION INERTIAL FUSION

In this section key issues in HIF are summarized first. The fuel target design should be conducted further toward a robust fuel implosion, ignition and burning. The HIF target design is quite different from the laser fusion target due to the relatively long range (the order of ~1mm) of the energy deposition [1, 17, 21].

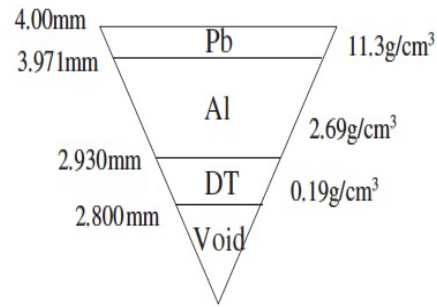

Fig. 1 An example fuel target structure in heavy ion inertial fusion.

An example direct-drive fuel target is presented in Fig. 1. The target should be compressed to about one thousand of the solid density to reduce the driver energy and to enhance the fusion reactions [22]. The target should be robust against the small non-uniformities caused by the driver beams' illumination non-uniformity, a fuel target alignment error in a fusion reactor, the target fabrication defect, et al. The ICF reactor operation would be 10~15 or so. So the stable target performance should be realized.

The HIB stopping range is rather long, and the HIB beam energy is deposited mainly at the end area of the beam ion stopping range due to the Bragg peak effect, which is originated from the nature of the Coulomb collision. The interaction of the HIB ions could be utilized to enhance the HIB preferable characteristics. The HIB ion interaction is relatively simple, and is almost the classical Coulomb collision, except the plasma range-shortening effect [6, 23]. Therefore, the HIB energy deposition profile is well defined in the HIF target. A typical HIB ion species would be $Pb^+$ or $Cs^+$ or so. For $Pb^+$ ions the appropriate Pb ion energy would be about 8GeV or so.

However, the HIBs illumination scheme should be studied intensively to realize a uniform energy deposition in a HIF target. The ICF target implosion uniformity must be less than a few % [16, 17]. The uniformity requirement must be fulfilled to release the fusion energy. The multiple HIBs should illuminate the HIF target with a highly uniform scheme

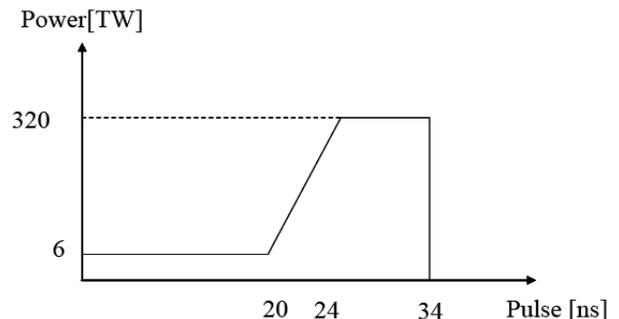

Fig. 2 An input heavy ion beam pulse. The HIB pulse consists of the low power part (foot pulse) and the high power one (main pulse).

during the imploding DT shell acceleration phase.

In addition, the HIB pulse shape should be also designed to obtain a high implosion efficiency $\eta_{imp}$. An example Pb HIBs pulse shape is presented in Fig. 2. The pulse shape consists of a low-intensity foot pulse, a ramping part to the peak intensity and the high-intensity main pulse. The foot pulse generates a weak shock wave in the target material and the DT fuel, and the first shock wave kicks the low-temperature DT liquid fuel inward. When no foot pulse is used, the main pulse with the high intensity generates a strong first shock wave inside of the DT fuel and increases the DT adiabat. The DT fuel preheat would be induced and the efficient fuel compression would be realized. The foot pulse length and the ramping time are designed to reduce the entropy increase. The first weak shock wave is not caught by the second and third stronger shocks inside of the DT fuel layer. At the inner edge of the DT fuel layer the shocks should be overlapped so that the efficient fuel acceleration and compression are accomplished during the implosion. A stronger shock wave increases the entropy more from the fluid ordered motion than the weak shock wave. The detail pulse shape should be designed for each target design.

The DT fuel stable compression is another key issue in ICF. In the recent NIF target experiments in Livermore, CA, U. S. A., they have succeeded to compress the DT fuel to a few thousands of the solid density and reached a scientific breakeven in ICF [24, 25]. Before getting the success on NIF, the low-intensity foot pulse lasers were used, and the DT fuel was compressed more than that shown in Refs. [24, 25]. The highly compressed DT fuel met a DT fuel mix, that would be induced by the R-T instability. The R-T instability and the fuel mix are caused by the implosion non-uniformity, the driver laser illumination non-uniformity, etc. The DT fuel implosion stability should be studied carefully in ICF. The DT fuel compression dynamics itself in HIF is the same as that in the laser fusion, though the driver target interaction is quite different from that in the laser fusion.

Each target should be injected into the reactor, and must be situated at the reactor center. The target alignment or positioning error in the reactor should be minimized to reduce the HIBs illumination non-uniformity. The experimental results show that the target alignment error would be minimized to ~100μm or so [26]. This issue is also studied for a real reactor design. On the other hand, the target should be designed to be robust against the target alignment of ~ 100μm.

The HIF reactor would have a dilute reactor gas inside the reactor chamber after each fuel target shot. The reactor operation frequency would be 10~15 Hz in HIF. In HIF the debris gas would supply cold electrons to compensate the HIB self charge in the vicinity of the fuel target and / or during the HIBs transport in the reactor. The reactor radius may be 3-5m or so. The debris gas dynamics should be also studied.

The HIB transport in the reactor is another key issue. The HIB ion mass is large so that the ion trajectory is almost straight between the HIB accelerator exit to the reactor center. However, each HIB carries a large current of ~1kA or so. The remaining large current and its self-charge would provide a slight HIB radial expansion [18]. The final neutralized HIB transport should be also studied.

The reactor design is also another key issue in ICF [2-4]. The first wall could be a wet wall with a molten salt or so or a dry wall. The reactor design must accommodate a large number of HIBs beam port, for example, 32 beam ports. At the first wall and the outer reactor vessel the beam port holes should have mechanical shutters or so to prevent the fusion debris exhaust gas toward the accelerator upstream. In addition, the target debris remains inside of the first wall or mixes with the liquid molten salt, which may be circulated. The target debris treatment should be also studied as a part of the reactor design. The tritium (T) also remains inside of each target after its burning. Usually about 30% or so of the DT fuel is reacted and depleted in the target during the burning process. So a large part of T in each target is mixed in the reactor gas and would be melted in the liquid first wall. The rest T and the radio-activated target materials must be distributed inside of the reactor vessel right after the target burning, and they would not accrete in a large lump in the reactor. The distributed radioactive materials must be collected and separated in the fusion reactor system safely.

The HIB accelerators have the high controllability and flexibility for the particle energy, the beam pulse shape, the pulse length, the ion species, the beam current, the beam radius, the beam focusing and the beam axis motion. The accelerators are operated daily with a high frequency stably. This is a very preferable feature for the fusion reactor system [27]. However, the high-current operation may need additional studies for the fusion reactor design. The ~kA HIB may induce an additional HIB divergence, a beam loss and an electron cloud generation in the accelerator. The high-current and high-charge HIB generation and transport should be studied carefully to avoid the uncertainty in the HIB accelerator. At present an experimental device of NDCX II at Berkeley, CA, U. S. A. works on the HIB accelerator physics study [28]. FAIR (Facility for Antiprotons and Ion Research) at Darmstadt, Germany and HIAF (High Intensity heavy ion Accelerator Facility) in China have been planned for HIF and HEDP (High Energy Density Physics) studies [29, 30]. The key issues in the accelerator physics include the ion source, which should supply a large number of HIB particles with 10-15 Hz or so [31]. The high-current HIB accelerator type should be also another issue in HIF accelerators [32].

In the accelerator system, each beam is generated first from the ion source, and near the final beam transport to the HIF reactor each HIB may be compressed longitudinally. The HIB pulse shape in Fig. 2 may impose a requirement of the HIB longitudinal length. For example, an 8GeV Pb ion has the speed of ~0.29c, where c is the speed of light. The 20ns HIB pulse length is about 1.7m. So at the last stage of the accelerator, the HIB may be compressed or bunched in the longitudinal direction. The HIB longitudinal bunching may be also required.

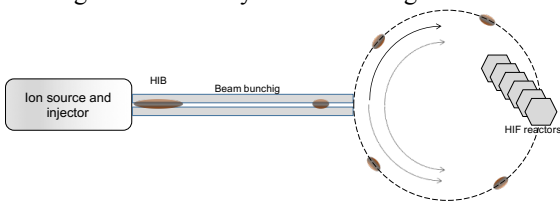

Fig. 3 Schematic figure for a HIF concept.

In addition to the issues shown above, the following pints should be studied for a realistic HIB reactor system: a power flow in the HIB power plant has not been well studied [33]. The ICF reactor operation is intermittent, and the DT neutrons and the target debris transfer the fusion pulse energy to the heat recovery system. The pulsation of the heat transfer would not have a significant influence on the steady electric power generation. This point should be studied to ensure the optimistic presumption, as well as a reactor design and its system design.

## 3 TARGET GAIN REQUIREMENT FOR HIF REACTOR SYSTEM

A target energy gain required for an energy production is evaluated by a reactor energy balance in ICF shown in Fig. 3. The driver pulses deliver an energy $E_d$ to a target, which releases fusion energy $E_{fusion}$. The energy gain is $G = E_{fusion}/E_d$. The fusion energy is first converted to electricity by a standard thermal cycle with an efficiency of $\eta_{th}$. A fraction $f$ of the electric power is circulated to the reactor system operation and the driver system, which converts it to the HIB energy with an efficiency of $\eta_d$. The energy balance for this cycle is written by $f\eta_{th}\eta_d G > 1$. Taking $\eta_{th} = 40$ % and requiring that the circulated-energy fraction $f$ of electrical energy should be less than 1/4, we find the condition $G\eta_d > 10$. For a driver efficiency in the range of $\eta_d = 30 \sim 40\%$, the condition $G > \sim 30$ is required for power production. Therefore, the preferable fusion target gain would be $G \sim 50\sim70$ in HIF. When the HIF reactor system operation is about 10~15 Hz, a 1GW HIF power plant can be designed.

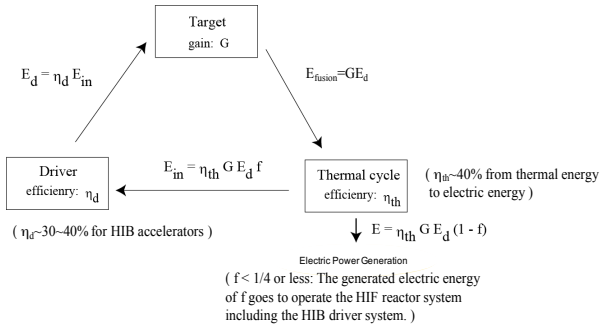

Fig. 4 Energy balance in ICF reactors.

## 4 HIB TRANSPROT IN A FUSION REACTOR

The HIB should be focused and transported against the beam space charge onto a fuel pellet at the reactor center. The target radius is the order of mm. In this section the HIB final transport is focused and studied. In the HIB final transport, the key factors are as follows: the final small focal radius (a few mm), a low emittance growth relating to the HIB particle energy and momentum divergences, the HIB space charge and current neutralizations, electrostatic and electromagnetic instabilities, and collision effects between the HIB and a background reactor gas [34, 35]. After the acceleration of beam ions to, for example, about 8-10 GeV, HIB should be transported in a fusion reactor of a few m to 5 m in the HIF reactor and focused on a few mm fuel pellet (see Fig. 1) [5, 21]. In a reactor, HIB particles interact with each other and with background plasma or a reactor gas (~ few mTorr) [36]. It was confirmed by D.A.Callahan that the interaction between neighboring HIBs is not serious in a HIB fusion reactor [37]. In the long distance transport of HIB, the HIB space charge and current should be neutralized, and electrostatic and electromagnetic instabilities should be suppressed. Consequently, we should establish a charged-particle-beam-transport control method for the above purposes.

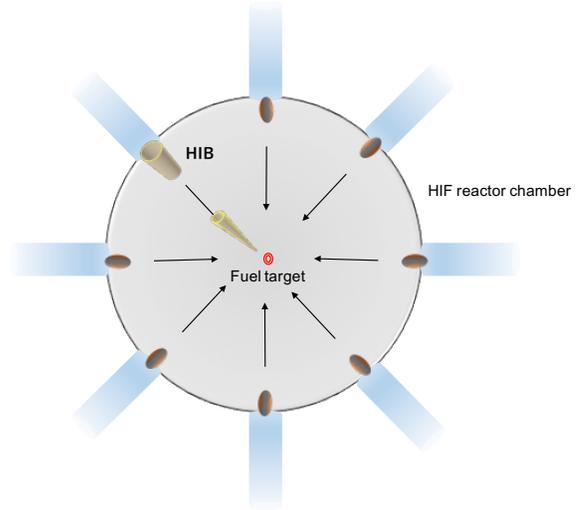

Fig. 5 Conceptual diagram for HIF reactor chamber

In the HIB final transport, several transport schemes have been proposed: the preformed channel transport [38, 39], the neutralized ballistic transport using a preformed plasma [38, 40, 41] or using a tube liner [36, 42], the ballistic transport in near vacuum [38], and so on. One of the promising transport schemes is the neutralized ballistic transport, in which preformed-plasma electrons or wall-emitted electrons neutralize the HIB space charge. On the other hand, the HIB ion number density increases from $n_{b0} \sim 10^{11} - 10^{12} \text{cm}^{-3}$ at a beam port entrance to $100 \sim 200 \times n_{b0}$ at the fuel pellet position. During the HIB transport in a fusion reactor, the HIB radius changes from several cm to 2–3mm. A chamber gas or attached electrons neutralize the HIB space charge well at the beginning transport section near the reactor wall. Near the fuel pellet at the chamber center, the HIB number density and the neutralizing electron number density increase, and we can also expect the chamber gas photoionization to increase the chamber gas electrons near the target. However, at the transport middle stage between these two regions, there may be a region in which the background chamber electron number density is smaller than the HIB neutralizing electron number density, depending on the chamber gas density and its ionization degree.

When a core plasma is surrounded by a lower density plasma or vacuum and at the same time the core plasma electron has a high temperature, the core plasma expands with a higher speed than the plasma ion thermal speed: the high-temperature electrons move out of and back to the core plasma ions and induce a charge separation at the core plasma ion surface. The charge separation extracts the plasma ions and, consequently, the core plasma expands fast by the ambipolar (plasma sheath) field. In the HIB transport, the HIB ions are the core plasma ions, and the neutralizing electrons may have a high temperature during the HIB convergence. Hereafter, we discuss the HIB divergence by the ambipolar field generated by the neutralizing electrons, propose possible solutions to suppress the HIB divergence and present a transportable window.

### 4.1 Neutralizing electron heating effect on HIB final transport near the chamber center

The fuel target requires several MJ of HIB driver energy in HIF. Each HIB carries 1–5 kA and a HIB ion particle energy may be 8–10GeV, depending on the HIB ion species. At the beam port entrance at the reactor wall a $Pb^+$ HIB radius $r_{b0}$ may be several cm and its number density is $n_{b0} \sim 10^{11}$–$10^{12}$ at the beam port entrance. The $Pb^+$ HIB radius decreases to $r_b \sim 2$–3mm in its radius at the target surface and its number density increases to $n_b \sim 100 \sim 200 \times n_{b0} = 1 \sim 2 \times 10^{14}$ at the fuel pellet position. In this section, we focus on the neutralized ballistic transport (NBT) scheme. In this scheme, a chamber gas pressure may be a few torr and a chamber neutral gas density may be $10^{14}/cm^3$. The chamber gas electron number density may be $10^{12}/cm^3$. In the NBT scheme, electrons are supplied from a preformed plasma or a plasma at a wall, and move together with a HIB to neutralize the HIB space charge. Along with a HIB convergence, the co-moving electron density increases to the same order of the HIB density. Near the chamber wall, the co-moving electron temperature $T_{e0}$ would be several tens of eV, and near the target area the electron temperature increases to a few hundreds of keV with the decrease in the HIB radius [41, 42]: if we assume an adiabatic increase in the electron temperature $T_e$, $T_e \sim T_{e0} \times (r_{b0}/r_b)^{4/3} \sim 22 \times T_{e0}$.

If we have a high number density $n_{ce}$ ($>Z_b n_b$) of the chamber electrons surrounding the HIB compared with the HIB ion number density $n_b$, like a channel transport scheme, the HIB space charge is always well neutralized [43]. At the central target area surrounding a fuel pellet in the reactor chamber, photoionized electrons are also supplied from the chamber gas, the HIB is also ionized to $Z_b \sim 6$ [43,44], and one can expect the situation of $n_{ce} > Z_b n_b$. The photoionization effect helps additionally neutralize the HIB charge at the central area of the reactor chamber. Therefore, at the central area the HIB can be expected to be well neutralized. Especially the additional electrons and the chamber original gas electrons are relatively cold than that of the electrons compressed at the chamber central area and co-moving with the HIBs from the reactor outer area.

On the other hand, near the chamber wall area, the HIB number density $n_{b0}$ would be comparable to the background chamber gas electron number density. At the same time, the emitted electrons move together and neutralize the HIB space charge well.

However, at the middle area in the NBT between the chamber wall area and the target central area, there may be a situation in which $n_{ce} < Z_b n_b$ and the neutralizing electron temperature $T_e$ becomes high. In this region a charge separation between the neutralizing electrons and the HIB ions induces a strong radial electric field to expand the HIB radius (see Fig.6). The ion acceleration mechanism is the same with that in the TNSA (target normal sheath acceleration) in laser plasma interaction [45-47].

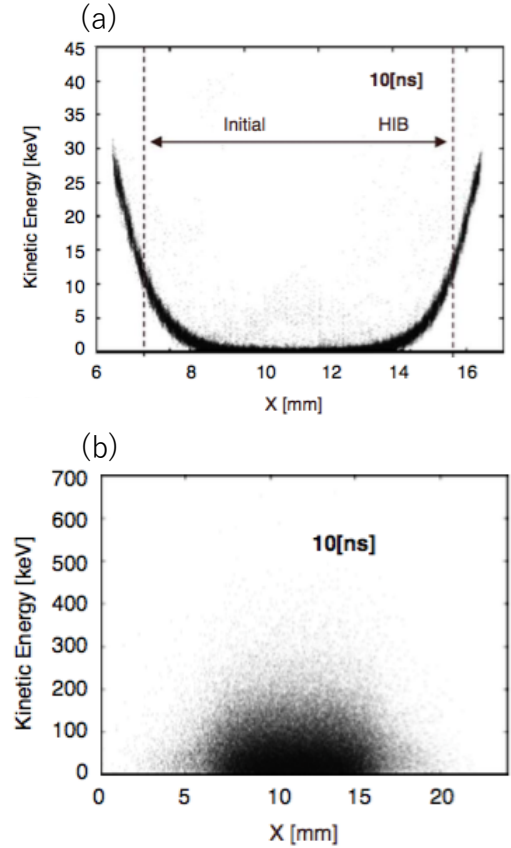

Fig. 7 PIC simulation results for the HIB transverse expansion near the reactor chamber center by the ambipolar field, which is generated by the electron heating. (a) The Pb ion transverse kinetic energy distribution in $x$, and (b) the electron energy distribution in $x$. In this simulation $T_e$= 100keV, $T_b$= 1keV, $r_b$=0.4cm, $n_b$=$10^{13}cm^{-3}$, $T_i$=1keV and $Z_b$=3

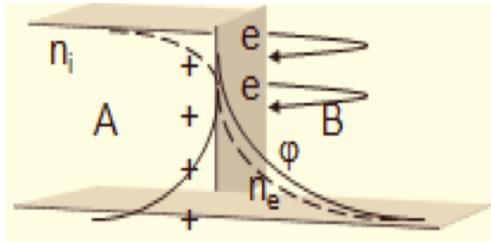

Fig. 6. Neutralizing high-temperature electrons compressed during the HIB focusing generate a charge separation at the HIB surface in the HIB transverse direction. The ambipolar field contributes to a slight expansion of the HIB in the radial direction.

In NBT, if the neutralizing electron number density is $n_{be}=Z_b n_b=10^{14}$ cm$^{-3}$, the neutralizing electron temperature is $T_e \sim 100$ keV and $n_{ce} < Z_b n_b$, the Debye length of the neutralizing electrons is $\lambda_{De} \sim 0.235$mm ($<r_b$). In the region A (inside of HIB) in Fig. 6, we can assume $n_{be} \sim Z_b n_b$ and in the region $B$ the ambipolar field potential can be estimated by the following Poisson equation:

$$\frac{d^2\varphi}{dr^2} \sim 4\pi e n_{be} \sim 4\pi e n_{e0} e^{e\varphi/T_e} \quad (1)$$

In the region $B$, the nonlinear Eq. (1) has an exact solution:

$$e\varphi = T_e \left\{ 1 - 2\ln\left(1 + \sqrt{\frac{e}{2}} k_{De} r\right) \right\},$$

$$Z_b e E = Z_b T_e k_{De} \left\{ \frac{2\sqrt{e/2}}{1 + \sqrt{e/2}(k_{De} r)} \right\} \quad \text{for } r > r_b. \quad (2)$$

Here $k_{De} = 1/\lambda_{De}$. After integration of Eq. (2) between the HIB surface $r_b$ to $r_b+\lambda_{De}$, we obtain the energy gained at the HIB surface by the ambipolar field in the transverse direction. When $Z_b=5$, $T_e=200$keV, $r_b=0.4$cm and $n_b = 3\times10^{13}$cm$^{-3}$, the HIB Pb ions obtain $\varepsilon_\perp = 136$keV of the transverse energy. In this case, a 10 GeV Pb HIB expands by 3.7 mm in its radius by the ambipolar field after the 1m transport. For $Z_b=2$, $T_e=10$keV, $r_b=0.5$cm and $n_b=2\times10^{12}$cm$^{-3}$, $\varepsilon_\perp =3$keV. In this case, the 10 GeV HIB Pb ion beam expands by 0.5 mm after the 1m transport. For $\lambda_{De} < r_b$, which is normally satisfied in NBT,

$$\varepsilon_\perp \propto \sqrt{\frac{Z_b}{n_{bi}}} \frac{T_e^{\frac{3}{2}}}{r_b}. \quad (3)$$

Then, $\varepsilon_\perp / \varepsilon_\parallel = (3 \sim 10 \text{ keV})/(8 \sim 10 \text{ GeV})$. The key factor for the ambipolar expansion of the transverse expansion near the fuel target area comes from the electron temperature $T_e$ increase with the HIB focusing. The next important factor would be $Z_b$, that is, the HIB ion charge stripping effect. The possible solutions to suppress the ambipolar field are presented below. The ambipolar field supplies a new mechanism for the HIB divergence, in addition to the effects of the beam space charge, the HIB emittance, the chromatic aberration and the HIB aiming error [37, 38].

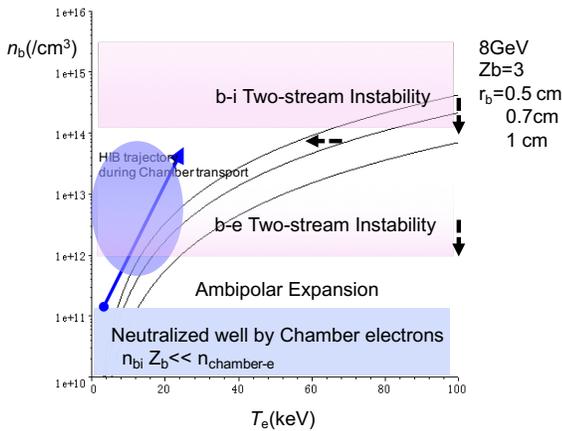

Fig. 8 HIB transport window in the neutralized ballistic transport (NBT) scheme in a HIF reactor chamber. The circled area is the transport window. The neutralizing electron temperature $T_e$ is limited by the ambipolar expansion, and the HIB ion number density is limited by two-stream instabilities. The solid arrow shows a preferable HIB trajectory during the drift compression and focusing in the reactor chamber.

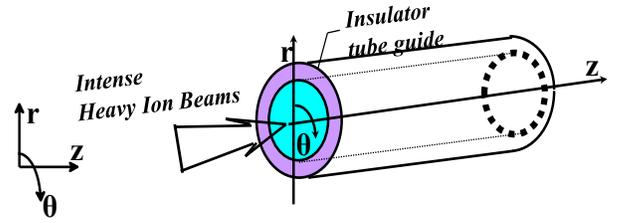

Fig. 9 Concept of an annular insulator guide for HIB charge neutralization. At the inner surface of the insulator guide, the high-current HIB creates a strong electric field and induces electric discharges, which create a plasma. From the plasma the sufficient electrons are extracted to neutralize the HIB space charge in the self-regulated manner.

The ambipolar HIB ion divergence phenomenon can be seen in Refs. [42-44], though the ambipolar expansion effect was not recognized in these references. In Refs. [43, 44], LSP simulations were performed, and ions located at the HIB outer part in transverse expand seriously. The co-moving electron temperature is also shown in Ref. [44] and shows a very high temperature of several tens to hundreds of keV. In Ref. [42], we have also shown the serious electron temperature increase as a HIB converges. In these simulation results, the effect of the ambipolar expansion was neither recognized nor pointed out. We clarified the HIB additional ambipolar expansion mechanism to explain the HIB divergence.

PIC simulations are also performed for the following parameter values for a Pb$^{3+}$ HIB: $T_e$=10 or 100keV, $T_b$= 1keV, $r_b$=0.4cm, $n_b$=10$^{13}$cm$^{-3}$, $T_i$=1keV and $Z_b$=3. Figure 7 shows the example results of (a) the Pb HIB particle kinetic energy, and (b) the electron energy distribution in $x$ for $T_e$=100 keV. The HIB, which is neutralized the background electrons, is located at the center of the simulation box in vacuum in this case. After 10ns, the HIB ions are accelerated transversely and obtain about $\varepsilon_\perp = 30$keV at the maximum. For $T_e$=10 keV, the maximal $\varepsilon_\perp$ is about 10 keV at 10 ns. This transverse energy $\varepsilon_\perp$ is another issue to obtain a fine focus on a fuel pellet.

When the chamber gas electron density $n_{ce}$ is much higher than the HIB number density, that is, $n_{ce} > Z_b n_b$, Eq. (2) also shows the HIB usual neutralization by the background chamber electrons. In this case, normally the chamber electron temperature is low. When $T_e$ =10 eV, $Z_b$=5, $r_b$=0.4 cm and $n_{ce}$ = 10$^{14}$ /cm$^3$, the HIB Pb ions obtain a negligible transverse energy of $\varepsilon_\perp$=0.05 keV. The neutralization by the high-density cold electrons is rather preferable for the HIB final transport

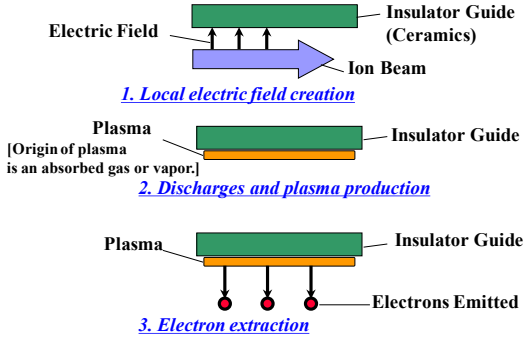

Fig. 10 Electron extraction mechanism to compensate the HIB space charge. An intense HIB creates a strong electric field at the insulator ceramics annular guide inner surface, and induces the electric discharges, which produces a plasma at the annular insulator surface. From the created plasma the HIB extract electrons to neutralize the HIB net charge in the regulated way.

in the reactor chamber.

### 4.2 Stable HIB transport window

The two-stream instability between the HIB ions and the background chamber electrons is excited as shown in Fig. 8, in which the lower limit band for the instability is indicated. However, this mode is an electron mode [48], and may not be so dangerous for the HIB transport, though it should be studied in detail. On the other hand, the two-stream instability between the HIB ions and the chamber gas ion also appears, but its limitation of $n_b$ is not serious. In Fig. 8 the solid arrow shows a possible preferable HIB transport trajectory in the reactor chamber (see Fig. 5). A circled region in Fig. 8 shows the transport window. The upper limit of the HIB ion number density $n_b$ comes from the two-stream instabilities [48], and the neutralizing electron temperature is limited by the ambipolar expansion limit driven by Eq. (2). For the ambipolar expansion limit we employ a relation of $\varepsilon_\perp/\varepsilon_\parallel = 10^{-6}$: for example, under this condition, the 50% increase in the HIB radius is allowed during 1m transport. In Fig. 8, three solid curves show the results from this limitation. The most upper solid curve is obtained under the conditions of $Z_b=3$, $r_b=0.5$cm and $\varepsilon_b=8$GeV. The middle solid curve is obtained for $Z_b=3$, $r_b=0.7$cm and $\varepsilon_b=8$GeV. The lower limit is for $Z_b=3$, $r_b=1.0$cm and $\varepsilon_b=8$GeV. For the low $n_b$ region ($n_{ce} > Z_b n_b$) no ambipolar field appears as we discussed above. Figure 8 demonstrates that the neutralizing electron temperature should be kept lower than the limit obtained by Eq. (2). This may be realized by a careful control of electron temperature, and the neutralizing electron supply method should be studied further [44].

If the background cold electron number density $n_{ce}$ is nearly equal to the HIB neutralizing high-temperature electron number density $Z_b n_b$ hot, the effective temperature $T_e$ becomes $T_{\text{eff}} = 1/(1/T_{ehot}+1/T_{ecold})$ in Eq. (3). Therefore, the ambipolar field is weakened by employing a cold background gas surrounding the HIB with a high density of $n_{ce} \geq Z_b n_b$.

Based on the analyses and discussions in this section, one can find the following two solutions in order to suppress

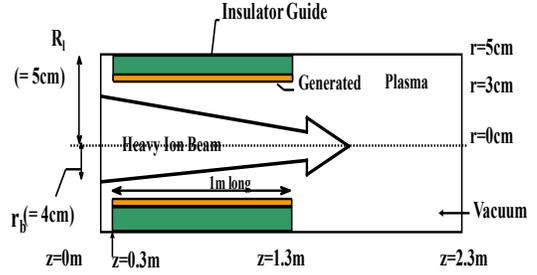

Fig. 11 HIB transport system using an annular ceramics insulator beam guide. The intense HIB creates a local electric field on the inner surface of the annular insulator guide. The local electric field induces the local discharges, and the plasma is produced on the insulator inner surface. The electrons are extracted from the plasma. The HIB space charge and current are effectively neutralized.

the ambipolar field divergence of HIB and to realize the stable HIB transport in the fusion reactor chamber:

(1) Use of a high chamber background electron density, which should be always larger than or equal to the HIB number density; this means $n_{ce} > Z_b n_b$. This means that near the target $n_{ce} > 10^{14}$–$10^{15}$/cm$^3$ and at the middle transport area in the reactor chamber $n_{ce} > 10^{13}$–$10^{14}$/cm$^3$. In this case the HIB is secure from the ambipolar field expansion, and the background electrons always come around the HIB ions to compensate the HIB space charge. This solution is simple, but in this case the HIB-chamber gas instabilities, the HIB charge stripping effect, and an interaction between a shock wave generated by a fuel pellet explosion and a chamber gas should be studied further in this circumstance. However, relating studies have been intensively performed in association with a channel transport scheme [37, 38, 49], and we can adopt the previous research results [50, 51].

(2) Employing the NBT scheme: in this case the lower temperature electrons should be used to neutralize the HIB space charge, in order to reduce the ambipolar field strength, based on Eqs. (2) and (3) near the chamber center, where the HIB is compressed in the radial direction.

Here it should be pointed out that the co-moving electron interaction with the background chamber gas is negligible. For the background electron temperature of 10eV and its density of a few Torr, a few eV is the co-moving electron collision energy loss. So the co-moving electrons survives and neutralize continuously during the whole HIB transport in the reactor chamber, once the HIB catches the neutralized electrons near the reactor wall chamber or just after the focusing element in the accelerator. From this estimation, after the HIB and co-moving electrons enter the HIB ICF reactor chamber gas, the electrons keep on moving with the HIB, and the HIB space charge neutralization is kept.

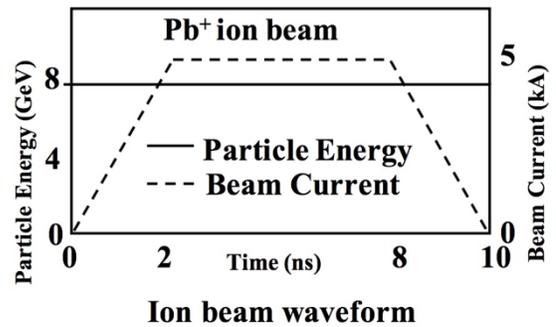

Fig. 12 Input Pb$^+$ ion beam waveform. The Pb ion-beam-parameter values are as follows: the maximum current is 5 kA, the particle energy is 8 GeV, the pulse width is 10 ns, and the rise and fall times are 2.0 ns.

Here we briefly review the two-stream and the filamentation instabilities [48, 52-55] in the reactor chamber, discussed in Fig. 8. The maximum growth rate of the two-stream instability between the HIB ions and the background electrons is given by Eq. (4).

$$\gamma_{max} = -\frac{\nu}{2} + \sqrt{\frac{\pi}{2} \frac{\omega_b^2}{\omega_e} \frac{V_b^2}{u_b^2}} \exp\left(-\frac{1}{2}\right) \quad (4)$$

Here the collision frequency is evaluated by the Coulomb interaction between the background electron and the HIB ions, whose speed is $V_b$ and thermal speed is $u_b$. The HIB ion plasma frequency is denoted by $\omega_b$, and $\omega_e$ is the background electron plasma frequency.

The maximum growth rate of the filamentation

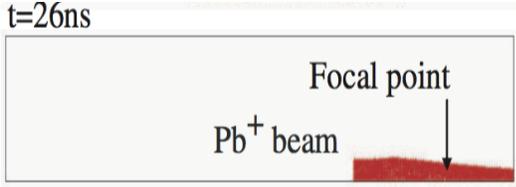

Fig. 13 Pb$^+$ ion particle map in the case for the ballistic focusing without the insulator guide. The mean velocity of Pb ions is given to focus at Z=2.1m. The radius at the focal point is about 5.8 mm, which is too large compared with the expected focal radius of about a few mm.

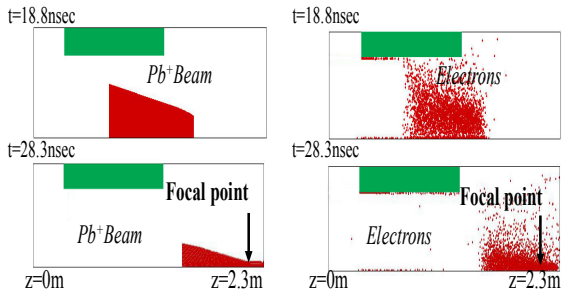

Fig. 14 Pb ion particle map and electron map in the case with the annular insulator guide. The focal radius is 2.4 mm at Z=2.1 m.

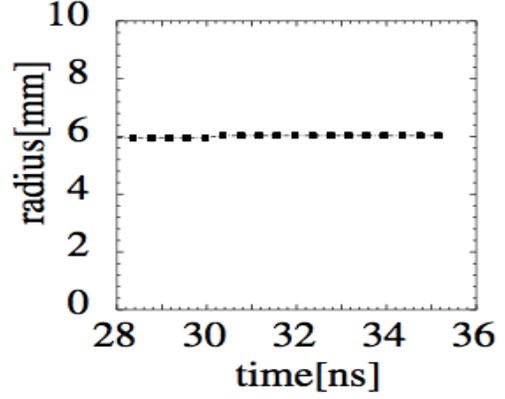

Fig. 15 HIB radius at the focal spot without the insulator ceramics guide. Due to the HIB space charge the HIB is not well focused at the focal point.

instability is given by Eq. (5).

$$\gamma_{max} = 2 \frac{\omega_b^2}{\omega_e} \frac{V_b^2}{u_b^2} \nu \quad (5)$$

Here the beam effective temperature is estimated by the inward motion of the focusing beam ions. An outer part of the focusing beam moves inward and overlaps an inner part, so that the beam effective temperature becomes high (~ MeV).

In Fig. 8 we calculate the density range in which the focusing HIB can propagate without severe influences of the two-stream instability ($\gamma\tau < 5$) by Eq. (4). The typical parameter values in HIF are as follows: the focusing beam temperature $T_b$~0.3~0.5MeV, the background plasma temperature $T_e$~10 eV, the beam drift velocity $V_b \sim 5 \times 10^9$ cm/s for 8 GeV Pb, the focused beam density $n_b \sim 1.38\times10^{11}$ ~ $2.5\times10^{12}$ cm$^{-3}$, the beam current is I~5.0 kA, and the background plasma density $n_e \sim 10^{15}$~$10^{16}$ cm$^{-3}$. For the filamentation instability, $\gamma\tau \sim 0.0040$~$0.0005$, and therefore, we confirm that the HIB is safe from the filamentation instability. We expect that the HIB is not influenced seriously by both the instabilities, as long as the background reactor gas electron density is kept low as shown in Fig. 8.

We have also checked that the HIB is safe from instabilities of a mode between HIB ions and reactor gas ions and modes between co-moving electrons and reactor gas ions. The modes have almost no influences on the HIB propagation in the HIF reactor.

### 4.3 HIB charge Neutralization

In this subsection, the HIB self-charge neutralization method is focused. In a HIF reactor system a HIB accelerator and final focusing elements should stand away from a reactor vessel so as not to be damaged by neutrons and fusion debris. In order to ensure the HIB transport control and its fine focus on a fuel target after a long distance HIB transport in the HIF reactor chamber, the HIB charge neutralization is needed. To realize the HIB fine focusing on a fuel pellet, the space-charge neutralization of the incident focusing HIB is required at the HIB final transport after the final focusing element in an HIB accelerator. An active plasma neutralization was realized experimentally in Ref. [41] at Berkeley. The actively-supplied plasma supplies electrons to neutralize the HIB net charge well, and the HIB fine focus was realized.

As a practical HIB neutralization method in an HIF reactor system, an insulator annular tube guide is proposed at the final transport part, through which a HIB is transported. The physical mechanism of HIB charge neutralization based on an insulator annular guide is as follows (see Fig. 9) [36, 56-58]: a local electric field created by the intense HIB induces local discharges, and a plasma is produced at the annular insulator inner surface. The electrons are extracted from the plasma by the HIB net space charge. The electrons emitted neutralize the HIB space charge well, and move together with the HIB in the reactor chamber. In addition, the HIB self-charge regulates the electron extractions to compensate the HIB charge as shown in Fig. 10.

We study the HIB transport through the annular insulator beam guide by particle-in-cell (PIC) simulations [59-62] in order to neutralize the beam space charge just after the final focusing element in an HIB accelerator. The results obtained in this study present that the HIB space charge is neutralized well by using the annular insulator guide located at the final transport in an HIB accelerator. Without a neutralization mechanism at the final transport part, the HIB final fine focus onto a fuel pellet becomes difficult.

When the magnitude of the electric field is beyond the threshold at the guide surface, the electric discharges are induced by the electric field and the local discharges produce the plasma (see Fig. 10). We assume that the plasma consists of protons and electrons, and that the thickness of the plasma layer is infinitesimal. We also assume that a sufficient amount of plasma is generated at the insulator guide surface such that the charged particle emission from the insulator inner surface is limited by the space charge and the plasma temperature is 10 eV. Then, the electrons are extracted from the plasma generated on the inner surface of the insulator beam guide by the HIB net space charge. The emitted electrons follow the HIB, and the HIB space charge is effectively neutralized by the electrons. Therefore, the HIB can be transported efficiently and one may expect a fine focus through the insulator beam guide. In this subsection, we employ a Pb ion beam to demonstrate the viability of the proposed insulator beam guide system. Our simulation model is shown in Fig. 11. We assume that the phenomenon concerned is cylindrically symmetric (see Fig. 9). The PIC code used is a 2.5-dimensional electromagnetic one.

The Pb ion-beam-parameter values are as follows: The maximum current is 5 kA, the particle energy is 8 GeV, the pulse width is 10 ns, and the rise and fall times are 2.0 ns as shown in Fig. 12. The initial beam radius is 4 cm at $Z=0$. The initial mean velocity of the focusing Pb beam is given to focus at $Z=210$ cm, and the average speed of the HIB ions injected is determined by the waveform. The beam temperature is 10 eV and the beam ions enter uniformly at the beam entrance, that is, $Z=0$. The maximum beam density is $1.3\times10^{11}$/cm$^3$ at $Z=0$. The transport area is in vacuum initially. In our simulation, the local plasma is generated on the insulator guide surface, when the magnitude of the electric field exceeds the threshold for the local discharge. The threshold value employed in this work is $1.0\times10^7$V/m in this study. The most outer boundary of the simulation area is a conductor, and in actual situations or experiments [63], the current flows through the conductor to the insulator surface. The origin of plasma generated on the ceramics insulator inner surface is gas or vapor absorbed into the insulator ceramics surface. Therefore, the ceramics insulator itself can survive long enough [63].

First, we simulate a Pb$^+$ ion beam propagation in a vacuum without the insulator beam guide. The particle map of the Pb$^+$ beam ions is shown in Fig. 13. In this case, due to the beam space charge, the beam radius at $Z=210$ cm (the focal point) is about 5.8 mm, which is slightly larger than that of usual reactor-size fuel target and would not be acceptable in HIF. Figure 14 presents the particle maps of the HIB ions and the electrons emitted from the insulator beam guide for the case with the proposed insulator guide system with the same initial conditions. The electrons extracted from the plasma move along with the Pb ion beam. The emitted electrons neutralize the space charge of the beam ions effectively, and suppress the radial expansion of the beam. Figure 15 shows the history of the HIB radius at the focal point, and shows a rather large HIB radius due to the self space charge. Figure 16 shows the HIB radius history at the focal point with the insulator guide. The final focal radius is about 2.4 mm in the case with the insulator guide. Figure 17(a) shows the histories of the total space charges of the HIB ions and the electrons in the whole transport region. The beam space charge is neutralized rather well by the electrons emitted from the insulator beam guide. Figure 17(b) presents the total currents of the HIB and the electrons. The HIB current is also well neutralized by the co-moving electrons. Figures 18 show ( (a) and (b) ) the electron temperature and ( (c) and (d) )the HIB ion temperature in the $z$ and $r$ directions for the case with the insulator guide. Figures 18 (a) and (b) shows the electron

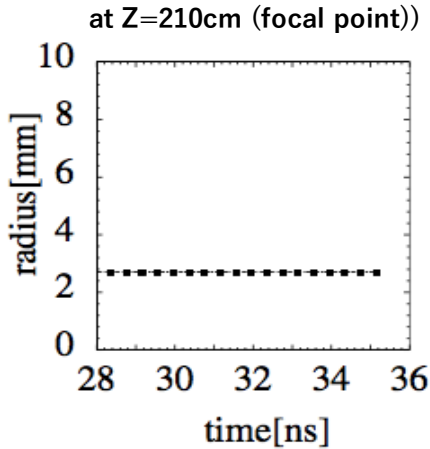

Fig. 16 HIB radius at $Z=210$cm (focal sport) with the insulator ceramics guide. The HIB space charge is well neutralized by the co-moving electrons extracted from the insulator guide inner surface.

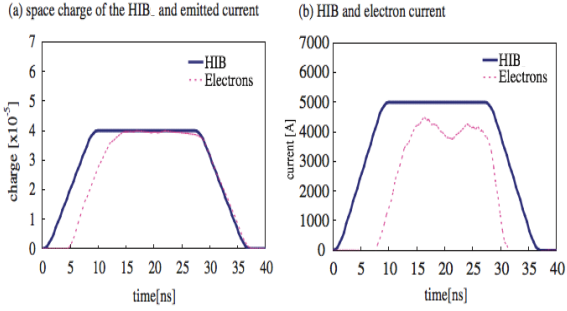

Fig. 17 (a) The total space charges of the HIB and the emitted electrons, which are extracted from the insulator ceramics guide inner surface and neutralize the HIB space charge well. (b) The HIB and the electron currents in the whole transport region.

temperature increase near the HIB focal point ($z$=210 cm). The electrons co-moving with the HIB are also focused with the HIB focusing, and are heated up to ~ a few hundreds keV, which induces the ambipolar field as shown in the previous subsection.

The HIB temperature also becomes large as it approaches the focal point, though the HIB temperature increase is not so significant, compared with the HIB ion longitudinal energy of ~8GeV. To find the HIB quality, we also evaluate the HIB emittance in the all-transport region for both cases: The emittance is 0.113 π-mm-mrad at $t$=0 ns, and 1.41 π-mm-mrad at $t$=24.3ns in the case with the insulator guide. On the other hand, the emittance is 2.07 π-mm-mrad at $t$=24.3ns in the case without the insulator guide. The results show that the HIB quality is kept well through the insulator guide.

## 5 PHYSICS OF FUEL TARGET IMPLOSION

In ICF, the DT fuel must be compressed to about ~1000 times of the solid density to reduce the input energy and increase the fusion reaction rate [22, 64]. The Lawson criterion shows the requirement for the density-radius product $\rho R$ of the compressed DT core based on the energy balance between the fusion energy output and the energy loss [22, 64]. In ICF the condition of $\rho R > $ ~0.5~1g/cm$^2$ should be fulfilled together with the DT hot core temperature > ~5~10keV. The DT burning fraction is estimated to be ~10~30% in the DT fuel target. A typical DT fuel target contains a few mg of DT in one target. So about several hundreds MJ of the fusion energy output is expected from each DT fuel target.

The fuel target implosion non-uniformity leads a degradation of fusion energy output. The implosion uniformity requirement is stringent. Therefore, it is essentially important to improve the fuel target implosion uniformity, and the implosion non-uniformity would be induced by the driver beam illumination non-uniformity and the other origins [14, 19, 65, 66]. The target implosion non-uniformity allowed is less than a few percent in inertial fusion target implosion [16, 17]. In this section first the non-uniform implosion effect is discussed on the target implosion in inertial fusion. Therefore, the driver energy deposition should be rather uniform to fulfill the uniform implosion requirement, and smoothing mechanisms are also expected to reduce the implosion non-uniformity. The R-T instabilities must take place at the ablation front and at the stagnation phase in the ICF target implosion. For the R-T instability at the stagnation phase, the perturbation amplitude and phase are defined by the non-uniform implosion process before the stagnation. Therefore, the uniform implosion at the acceleration phase is essentially important. The $\rho R$ product at the stagnation is directly relating to the fusion energy output. When a small implosion non-uniformity is imposed, $\rho R$ is degraded from the perfect uniform $(\rho R)_0$. $\rho R$ is proportional to $1/R^2$. Therefore, $(\rho R)/(\rho R)_0 = \{(R+\delta R)/R\}^{-2} = (1+\delta R/R)^{-2}$. On the other hand, the nonuniformity $\delta \alpha$ of the implosion acceleration $\alpha$ is estimated by $\delta\alpha/\alpha \simeq \delta R/r_0 = (\delta R/R)(R/r_0) = \eta^{-1/3}(\delta R/R)$, where $r_0$ is the fuel initial radius and $\eta$ the density compression ratio. Typically the density compression ratio $\eta$ is about 1000 in inertial fusion. Then we obtain the relation of $\delta\alpha/\alpha \simeq \eta^{-1/3}[\{(\rho R)_0/\rho R\}^{1/2} - 1]$. In an inertial fusion reactor the degradation threshold of $(\rho R)/(\rho R)_0$ would be about $0.5 \sim 0.8$, and $\delta\alpha/\alpha$ should be less than about 3~4% [17]. Based on this consideration, the driver beam illumination non-uniformity should be also mitigated to release a sufficient fusion energy output.

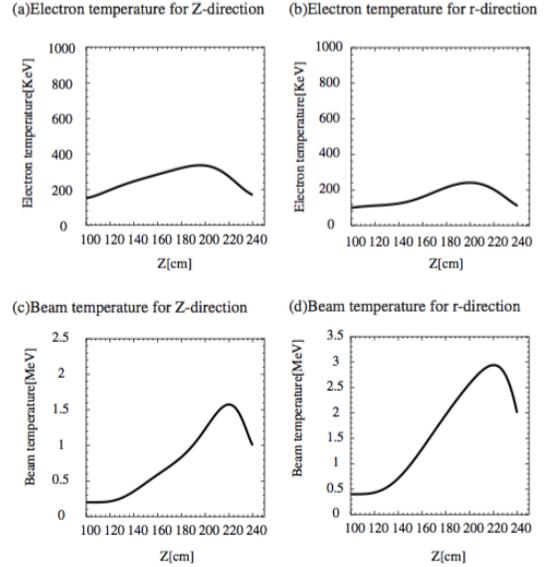

Fig. 18 The emitted electron temperature for (a) longitudinal and (b) trans- verse direction, respectively. The HIB temperature for (c) longitudinal and (d) transverse direction, respectively.

### 5.1 Target implosion dynamics in HIF

In order to study the fuel target implosion in HIF, we present a typical target hydrodynamics by using a 2D hydrodynamics code coupling with the HIB illumination code [20, 67]. We employ a 32-HIBs illumination system [68]. In this subsection, two-dimensional simulations are performed. The DT fuel target employed is shown in Fig. 1 and the HIB input pulse is shown in Fig. 2. In this specific case, the total HIB energy is 4.0 MJ.

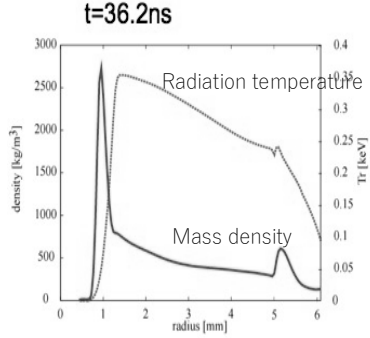

Fig. 19 A mean density and a mean radiation temperature averaged over the $\theta$ direction at 36.2 nsec.

Figure 19 presents a mean density and a mean radiation temperature averaged over the $\theta$ direction at 36.2 nsec. The averaged HIB illumination non-uniformity is 2.3 % in this case. Other 1D example simulations results are also found in Ref. [22], which presents some typical ion-beam-based DT fuel target implosion performances.

The HIB deposition energy distribution produces an ablation region at the Al energy absorber layer, and then about one-third of Al pusher mass pushes the DT fuel. In Fig. 19 the low density region appears at the ablation front. The density gradient scale length of the ablation surface is relatively large in HIF target implosion, that is, about several hundred μm~500μm or so.

When the density gradient scale length $L$ is large, the growth rate ($\gamma$) reduction effect on the R-T instability would be expected [14, 15]. $\gamma = \sqrt{gk/(1+kL)}$ Here $g$ is the implosion acceleration, and $k$ the wave number. In HIF, typically $L$ is about several hundred μm~500 μm as shown in Fig. 13 and Ref. [22], and the ablation effect is minor in HIF. Therefore, the short wavelength ($2\pi/k$) modes of the perturbation would be suppressed or mitigated by the density gradient effect in HIF. So in HIF typically the large scale perturbation modes, which have the wavelength of several hundred μm~500μm, becomes a central concern to keep the fuel target implosion spherically symmetric.

For the fuel dynamics near the ignition, a hot-spot dynamics analysis is also important at the final stage of the fuel stagnation. At the initial stage of the fuel target implosion the implosion non-uniformity reflects the imposed non-uniformity by, for example, the driver beam illumination non-uniformity, the target fabrication error and so on. During the implosion, the target radius shrinks till the fuel ignition, and the initial amplitude of the imposed non-uniformity grows as the fuel radius shrinks. However, the previous implosion simulation results [17, 67] presented that the non-uniformity grows significantly and the non-uniformity amplitude enhancement is extraordinary especially around the ignition or final stagnation phase.

When the fuel behavior near the ignition is treated by a fluid model in the Lagrangian form, the Kelvin's theorem shows conservation of the vorticity $\omega$ [69]: $\omega S = constant$. Here $S$ is the circulating area. In inertial fusion, the DT fuel mixing is induced by the R-T instability and also by the non-uniform implosion. The fuel mixing is one kind of the vortex, that is, the circulating motion. During the shrinkage of the DT fuel at the stagnation final stage just before the fuel ignition, $S$ must be reduced together with the shrinkage of the fuel radius scale length $L_{DT}$. Therefore, the vorciticy $\omega$ would be enhanced significantly at the final stage of the stagnation as follows: $\omega \sim \omega_0 (L_{DT0}/L_{DT})^2$. In addition, the fusion fuel target shrinks in the 3 dimensional way. So in the Lagrangian frame the fuel mass is conserved during the stagnation, and the Ertel theorem [69, 70] shows $\omega/\rho = \omega_0/\rho_0$. The ICF target fuel is compressed in 3D, and so $\omega \sim \omega_0 (L_{DT0}/L_{DT})^3$ (see Fig. 20). Based on this consideration, the vorticity $\omega$, that is, the circulation of the DT fuel would be enhanced significantly. The circulation enhancement induces the mixing of the cold fuel and hot fuel.

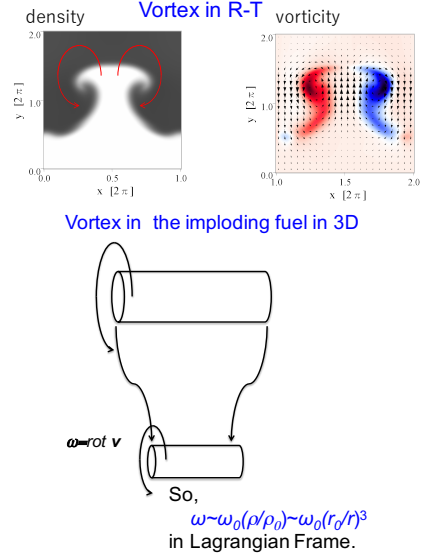

Fig. 20 The fuel mixing is one kind of the vortex, that is, the circulating motion. During the shrinkage of the DT fuel at the stagnation final stage just before the fuel ignition, the vorticity $\omega$, that is, the circulation of the DT fuel would be enhanced significantly.

Due to the fuel mixing enhancement at the fuel stagnation phase, the fuel non-uniformity would be significantly enhanced at the final stage of the fuel compression just before the fuel ignition. In order to release the DT fusion energy stably, the initial non-uniformity, which is the seed of the consequent fuel mixing at the final stagnation phase, must be reduced to a few % of the implosion non-uniformity.

### 5.2 Smoothing of HIBs illumination non-uniformity in HIF

In order to realize an effective implosion, beam illumination non-uniformity on a fuel target must be suppressed less than a few percent. In this subsection first a direct-indirect mixture implosion mode is discussed in heavy ion beam (HIB) inertial confinement fusion (HIF) in order to release sufficient fusion energy in a robust manner. In the direct-indirect mixture mode target, a low-density foam layer is inserted, and the radiation energy confinement is enhanced by the foam layer. In the foam layer the radiation transport is expected to smooth the HIB illumination non-uniformity in the lateral direction. Two-dimensional implosion simulations are performed [66, 67], and show that the HIB illumination non-uniformity is well smoothed in the direct-indirect mixture target.

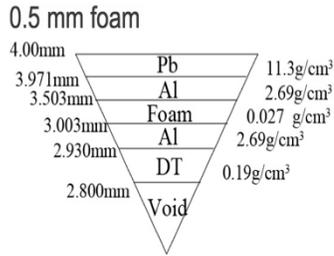

Fig. 21 Target structure with the 0.5 mm-thick foam. The foam inserted confines the radiation energy more to smooth the HIBs illumination non-uniformity.

Figure 21 shows a typical fuel target in HIF. The radiation energy confined may smooth the HIB illumination non-uniformity. Therefore, we employ a foam layer to increase the confined radiation energy at the low density region as shown in Fig. 19. We call this target as a direct-indirect hybrid target. The mass density of the foam layer is 0.01 times the Al solid density in this study.

The HIB pulse consists of a foot pulse and a main pulse as shown in Fig. 2. In this case, the total HIB energy is 4.0MJ. We employ the 32-HIBs illumination system [68]. We evaluate the beam illumination non-uniformity at the target. In HIF the Bragg peak deposition area plays the most important role for the fuel acceleration.

In the foam layer the radiation transport is expected to smooth the HIB illumination non-uniformity in the lateral direction. To see the radiation transport effect on the implosion non-uniformity smoothing, we compare the results for the cases with the radiation transport (ON) and without the radiation transport (OFF) for the target shown in Fig. 21.

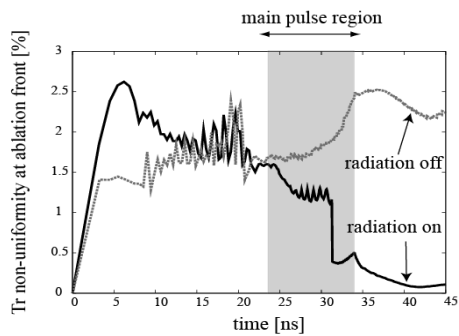

Fig. 22 The time histories of the RMS non-uniformity of the radiation temperature at the ablation front in the cases of the radiation transport ON and OFF.

Figure 22 presents the time histories of the RMS non-uniformity of the radiation temperature at the ablation front in the cases of the radiation transport ON and OFF. In Fig. 22 we see that the implosion non-uniformity at the ablation front becomes small effectively by the main pulse in the case of the radiation transport ON. During the main pulse, the implosion non-uniformity is well smoothed by the radiation transport effect.

In the direct-indirect hybrid implosion we employ the target with the 0.5mm-thickness foam as shown in Fig. 21. The peak conversion efficiencies of the HIB total energy to the radiation energy are ~ 4.5 % in the case of the 0.5mm foam and ~ 1.5 % in the case without the foam. The result means that a few hundreds kJ of the radiation energy is confined in the density valley and contributes to the non-uniformity mitigation. From these results, we find that the implosion mode in the case with the foam would be a direct-indirect hybrid mode.

Next a mitigation mechanism of the R-T instability is discussed in order to realize a spherically symmetric implosion. So far the dynamic stabilization for the R-T instability [71-76] has been discussed in order to obtain a uniform compression [22, 77] of a fusion fuel in ICF. The R-T dynamic stabilization was found many years ago [71, 72] and is important in inertial fusion. It was found that the oscillation amplitude of the driving acceleration should be sufficiently large to stabilize the R-T instability [71-76]. In inertial fusion, the fusion fuel compression is essentially important to reduce an input driver energy, and the implosion uniformity is one of critical issues to compress the fusion fuel stably. Therefore, the R-T instability stabilization [71-76] or mitigation [11-13] is attractive to minimize the fusion fuel mix.

On the other hand, instabilities grow from a perturbation in general, and normally the perturbation phase is unknown. Therefore, we cannot control the perturbation phase, and usually the instability growth rate is discussed. However, if the perturbation phase is controlled and known, we can find a way to control the instability growth. One of the most typical and well-known mechanisms is the feedback control in which the perturbation phase is detected and the perturbation growth is controlled or mitigated or stabilized. In plasmas it is difficult to detect the perturbation phase and amplitude. However, even in plasmas, if we can actively impose the perturbation phase by the driving energy source wobbling or so, and therefore, if we know the phase of the perturbations, the perturbation growth can be controlled in a similar way [11-13]. For example, the growth of the Weibel instability or the filamentation instability [48, 52, 53, 78] driven by a particle beam or a jet could be controlled by the beam axis oscillation or wobbling. The oscillating and modulated beam induces the initial perturbation and also could define the perturbation phase. Therefore, the successive phase-defined perturbations are superposed, and we can use this property to mitigate the instability growth. Another example can be found in heavy ion beam inertial fusion; the heavy ion accelerator could have a capability to provide a beam axis wobbling with a high frequency. The wobbling heavy ion beams also define the perturbation phase. This means that the perturbation phase is known, and so successively imposed perturbations are superposed on the plasma. We can use the capability to reduce the instability growth by the phase-controlled superposition of perturbations. Here we discuss and clarify the dynamic mitigation mechanism for instabilities. In HIF the HIBs illumination non-uniformity would be mitigated by the wobbling beam motion, that is, the HIB axis oscillation or rotation [11-13].

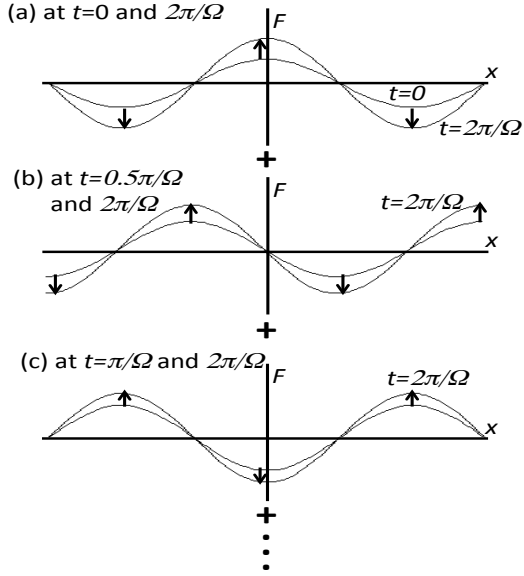

Fig. 23 Superposition of perturbations defined by the wobbling driver beam. At each time the wobbler provides a perturbation, whose amplitude and phase are defined by the wobbler itself. If the system is unstable, each perturbation is a source of instability. At a certain time the overall perturbation is the superposition of the growing perturbations. The superposed perturbation growth is mitigated by the beam wobbling motion.

In instabilities, one mode of an initial perturbation, from which an instability grows, may have the form of $a = a_0 e^{ikx+\gamma t}$, where $a_0$ is the amplitude, $k=2\pi/\lambda$ the wave number, $\lambda$ the wave length and $\gamma$ the growth rate of the instability. At $t$=0 the perturbation is imposed. The initial perturbation grows in an unstable system. After $\Delta t$, if the feedback control works on the system, another perturbation, which has an inverse phase with the detected amplitude at $t$=0, is actively imposed, so that the actual perturbation amplitude is mitigated very well (see Fig. 23). This is an ideal example for the instability mitigation.

In plasmas the perturbation phase and amplitude cannot be measured dynamically. However, by using a wobbling beam or an oscillating beam or a rotating beam or so, the initial perturbation is actively imposed so that the initial perturbation phase and amplitude are defined actively. In this case, the amplitude and phase of the unstable perturbation cannot be detected but can be defined by the input driver beam wobbling at least in the linear phase. In plasmas it is difficult to realize the perfect feedback control, but a part of its idea can be adopted to the instability mitigation. In actual, heavy ion beam accelerators can provide a controlled wobbling or oscillating beam with a high frequency [8-10].

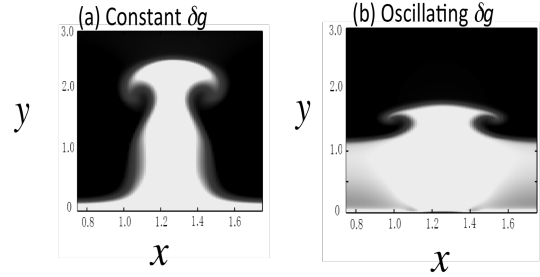

Fig. 24 Example simulation results for the R-T instability mitigation. (a) 10% acceleration non-uniformity drives the R-T instability as usual, and (b) 10% acceleration non-uniformity oscillates or wobbles.

If the energy driver beam wobbles uniformly in time, the imposed perturbation for a physical quantity of $F$ at $t=\tau$ may be written as $F = \delta F e^{i\Omega\tau} e^{\gamma(t-\tau)+i\vec{k}\cdot\vec{x}}$. Here $\delta F$ is the amplitude, $\Omega$ the wobbling or oscillation frequency, and $\Omega t$ the phase shift of the superposed perturbations. At each time $t=\tau$, the wobbler provides a new perturbation with the controlled phase shifted and amplitude defined by the driving wobbler itself. After the superposition of the perturbations, the overall perturbation is described as $\int_0^t d\tau\ \delta F e^{i\Omega\tau} e^{\gamma(t-\tau)+i\vec{k}\cdot\vec{x}} \propto \frac{\gamma+i\Omega}{\gamma^2+\Omega^2} \delta F e^{\gamma t} e^{i\vec{k}\cdot\vec{x}}$. At each time of $t=\tau$ the driving wobbler provides a new perturbation with the shifted phase. Then each perturbation grows by the factor of $e^{\gamma t}$. At $t>\tau$ the superposed overall perturbation growth is modified as shown above. When $\Omega >> \gamma$, the perturbation amplitude is reduced by the factor of $\gamma/\Omega$, compared with the pure instability growth ($\Omega$=0).

Figure 23 shows the perturbations decomposed, and each time the phase-defined perturbation is imposed actively by the driving wobbler. The perturbations are superposed at the time $t$. The wobbling trajectory is controlled by for example a beam accelerator or so, and the superposed perturbation phases and amplitudes are controlled so that the overall perturbation growth is controlled.

From the analytical expression for the physical quantity $F$, the mechanism proposed in this subsection does not work, when $\Omega \ll \gamma$. Only for the modes, which satisfy the condition of $\Omega \geq \gamma$, the mechanism of the instability mitigation by the wobbler is applied for its growth mitigation. For R-T instability, the growth rate $\gamma$ tends to become larger for short wavelengths. If $\Omega \ll \gamma$, the modes cannot be mitigated. In addition, if there are other sources of perturbations in the physical system and if the perturbation phase and amplitude are not controlled, this dynamic mitigation mechanism also does not work. For example, if the sphericity of an inertial fusion fuel target is degraded, the dynamic mitigation mechanism does not work. In this sense the dynamic mitigation mechanism is not almighty. Especially for a uniform compression of an inertial fusion fuel all the instability stabilization and mitigation mechanisms would contribute to release the fusion energy in stable.

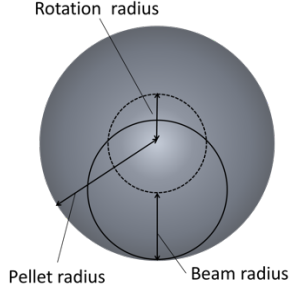

Fig. 25 Schematic diagram for a circularly wobbling beam

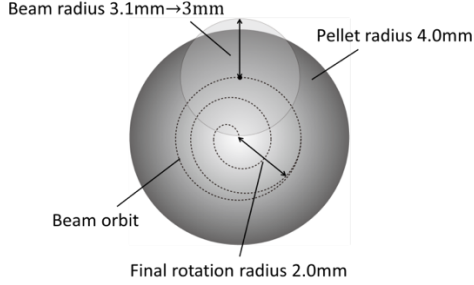

Fig. 26 Schematic diagram for spiral wobbling beam

Figures 24 show example simulations for the R-T instability, which has one mode. In this example, two stratified fluids are superposed under an acceleration of $g = g_0 + \delta g$. In this example case the wobbling frequency $\Omega$ is $2\pi\gamma$, the amplitude of $\delta g$ is $0.1g_0$, and the results shown in Figs. 24 are at $t = 8\gamma$. In Fig. 24(a) $\delta g$ is constant and drives the R-T instability as usual, and in Fig. 24(b) the phase of $\delta g$ is shifted or oscillated with the frequency of $\Omega$ as stated above for the dynamic mitigation. The example simulation results also support the effect of the dynamic mitigation mechanism well.

In HIF a fuel target is irradiated by HIBs, when the fuel target is injected and aligned at the center of the fusion reactor [19, 26, 42]. Here we employ (Pb+) ion HIBs with the mean energy of 8GeV. The HIB temperature is 100MeV and the HIB transverse distribution is the Gaussian profile. The beam radius at the entrance of a fusion reactor is 35mm and the radius of a fusion reactor is 3m. We employ an Al monolayer pellet target structure with a 4.0mm external radius. The 32-HIBs positions are given as presented in Ref. [68]. The HIBs illumination non-uniformity is evaluated by the global *rms*, including the Bragg peak effect in the energy deposition profile in the target radial direction. In this study, one HIB is divided into many beamlets, and the precise energy deposition is computed [19, 20, 26].

So far, we have found that the growth of the R-T instability would be mitigated well by a continuously vibrating non-uniformity acceleration field with a small amplitude compared with that of the averaged acceleration [11-13]. Figure 25 shows a schematic diagram for the wobbling beam. However, in our previous work [79] we found that at the initial stage of the wobbling HIBs illumination the illumination non-uniformity, that is, the initial imprint, becomes huge and cannot be accepted for a stable fuel target implosion.

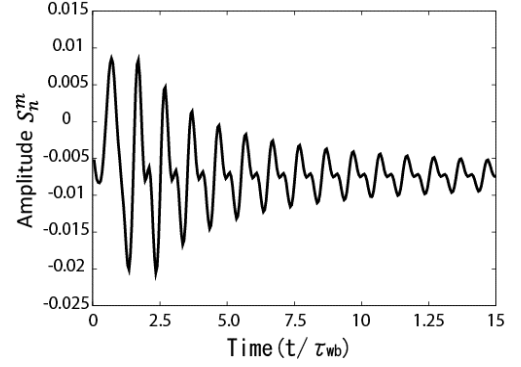

Fig. 27 The mode (2,0) amplitude of HIBs non-uniformity vs. time

This problem on the initial imprint of the rotating HIBs illumination is solved by the spiral wobbling HIBs as shown in Fig. 26. When the spirally wobbling beams in Fig. 26 are used, the initial imprint of the non-uniformity at the beginning of the irradiation is greatly reduced about from 14% to 4%. For the spiral wobbling beam the beam radius changes from 3.1mm to 3.0mm at $t = 1.3\tau_{wb}$. Here $\tau_{wb}$ is the time for one rotation of the wobbling beam axis. The beam rotation radius becomes 2.0mm at $t = 2.0\tau_{wb}$. After that, the beam rotation radius is 2.0mm. In our studies, we employ the spirally wobbling beam for the HIBs illumination non-uniformity study.

Figure 27 shows the amplitude of the mode $(n, m) = (2, 0)$ vs. time, and Fig. 28 presents the spectrum of the mode $(2, 0)$ in its frequency space. Here $(n, m)$ are the polar and azimuthal mode numbers, and $S_n^m$ is the amplitude of the spectrum, respectively. If the deposition energy distributed is perfectly spherically symmetric, the amplitude of the spectrum is 1.0 in the mode $(n, m) = (0, 0)$. For this reason, the amplitude of the mode $(n, m) = (0, 0)$ becomes large, nearly 1.0. In this case, the amplitude of the spectrum mode is not taken into consideration. As a result, the amplitude of spectrum mode $(n, m) = (2, 0)$ is largest and the mode $(n, m) = (2, 0)$ is dominant throughout the HIBs illumination. In Fig. 27 the time is normalized by the wobbling beam axis rotation time $\tau_{wb}$. In Fig. 28 $f_{wb}$ shows the wobbling HIBs rotation frequency. The result in Fig. 28 demonstrates that the small non-uniformity of the HIBs energy deposition has the oscillation with the same frequency and the double frequency with the wobbling HIBs oscillation frequency of $f_{wb}$.

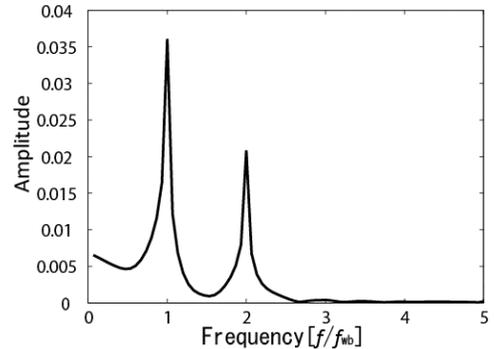

Fig. 28 Spectrum of the mode (2, 0) in its frequency space.

In addition, we have been investigating the influence of the wobbling HIBs energy deposition on the implosion acceleration in a HIF target implosion. Our preliminary results show that the implosion acceleration reflects well the wobbling HIBs energy deposition, and so the implosion acceleration $g$ has a small oscillating acceleration $\delta g$, which contributes to the R-T instability growth mitigation and the HIBs non-uniformity smoothing in HIF. The remarkable results come from the unique features of the HIB accelerator high controllability for the HIB's axis motion [8-13].

### 5.3 Tritium content of ICF DT fuel target

In ICF each DT fuel contains a few mg of DT fuel. In a second 10~15 fuel targets are ignited and burned. Therefore, each ICF reactor uses about ~1000kg of T in a year. The tritium has a half life of 12.3 years. The T production is not so difficult, and the reaction of Li + neutron is used to produce T in a fusion reactor. The total T inventory would be huge.

We have studied the T content in the DT fuel target. The results show that about ~30% reduction of tritium still produces a sufficient DT fusion energy output in a DT fuel target. This means that the T inventory can be reduced significantly without a significant reduction of the DT fusion energy output in an ICF reactor system. The important result comes from the contribution of the DD reactions in a DT fuel target and from a fact that a relatively small amount (~10~30%) of DT fuel is reacted in a DT fuel target.

In a DT fuel target the DT and DD reactions mainly contribute to the fusion reactions:

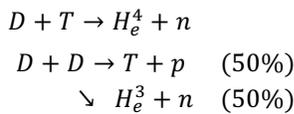

$$D + T \rightarrow H_e^4 + n$$
$$D + D \rightarrow T + p \quad (50\%)$$
$$\searrow H_e^3 + n \quad (50\%)$$

The contents of the important elements in a DT fuel target are estimated by the following simplified equations:

$$\frac{dn_T}{dt} \approx -n_D n_T <\sigma v>_{DT} + n_D^2 <\sigma v>_{DD}/4$$
$$\approx -n_T^2(-<\sigma v>_{DT} + <\sigma v>_{DD}/4) \quad (6)$$
$$\frac{dn_D}{dt} \approx -n_D n_T <\sigma v>_{DT} - n_D^2 <\sigma v>_{DD}$$
$$\approx -n_D^2(<\sigma v>_{DT} + <\sigma v>_{DD}) \quad (7)$$

Here $<\sigma v>$, which depends on only the temperature, is the reaction rate for each reaction and $t$ the time. $n_{T0} = n_0/2 - \delta, n_{T0} = n_0/2 - \delta$. Here we assume the temperature is constant and $\delta \ll n_0$. We obtain

$$\delta/(n_0/2) \sim (5/32)(n_0\tau)^2 <\sigma v>_{DT} <\sigma v>_{DD}/(1 + n_0\tau <\sigma v>_{DT}/2)^2, \quad (8)$$

for the optimal content of tritium in DT ICF. The result shows that $\delta$ is positive and the reduction of the T content brings sufficient fusion energy. For an example parameter set of T=100keV, $n_0 = 4.5 \times 10^{25}/cm^3$ and $\tau$=100ps, $\delta/(n_0/2) = 0.014$. This value means that 14% of the tritium reduction brings the maximal DT fusion energy output in the DT ICF. The additional numerical estimations and fluid computer simulations support the result as presented in Ref. [80]: the DT fusion output becomes peak around $\delta/(n\_0/2) \approx 0.01$, and that about 30% T reduction still gives a sufficient fusion energy output in an ICF reactor system.

## 6 HIB ILLUMINATION ON A FUEL TARGET

In HIF key issues includes a non-uniformity of heavy- ion-beam illumination onto a fuel pellet in a fusion reactor [19, 20, 68]. The HIB illumination non-uniformity must be suppressed less than a few percent in order to achieve a symmetric fuel pellet implosion. Previous researches [26, 81-83] show that in realistic cases a pellet displacement from a reactor chamber center may be required to be within about 20~100μm. If this requirement is relaxed, the target placement precision, the driver HIB alignment accuracy, and the target tracking accuracy may be also relaxed. Therefore, we study a robust HIB illumination scheme against the fuel pellet displacement. In this section, we study the robust HIB illumination scheme by three-dimensional HIB illumination simulations in a direct-driven HIB illumination scheme [19, 20], and found a robust HIB illumination scheme by using a larger radius of HIBs and by an optimization of HIB illumination angles. The new robust HIB illumination scheme allows a fuel pellet displacement of 100~200μm.

For the evaluation of the illumination non-uniformity on the target, we employ the *rms* non-uniformity on the target [84, 85], and also the mode analyses of the HIB deposition energy on the spherical fuel pellet using the Legendre-polynomial functions.

In our studies we employ $Pb^+$ ion HIBs with the mean particle energy of 8 GeV. The beam radius at the entrance of a reactor chamber wall $R_{en}$ is 35mm, the reactor chamber radius $R_{ch}$ is 5m. The beam particle density distribution is the Gaussian one. The longitudinal temperature of HIB ions is 100 MeV with the Maxwell distribution. The beam transverse emittance $\varepsilon_r$ is 3.5 mm mrad [67]. From the beam emittance $\varepsilon_r$, we obtain a divergence angle $\alpha_{dvr}$, the focal spot radius $R_f$, and $f$ (see Fig. 29) [20, 67]. $\varepsilon_r = (\pi/0.18)\alpha_{dvr}R_{en}$. The target temperature increases linearly during the time of a HIB pulse deposition from 0.025 to 300 eV in our study. We employ a Pb + Al pellet structure with the 4mm external radius. The Pb mass density is 11.3 g/cm$^3$. In the pellet structure, the outer Pb layer thickness is 0.06 mm.

In this study, we employ the 32-HIBs illumination system [68, 86]. Figure 30 presents a schematic image for the 32 HIB centers at the target surface. Figure 30 shows the HIB illumination displacement angle, which is used in this work to optimize the HIB illumination angle. When $\Delta\theta$ becomes zero, the HIB illumination scheme becomes the conventional one in Refs. 68 and 86. In this work we assume that all the HIB center lines are directed to a target center.

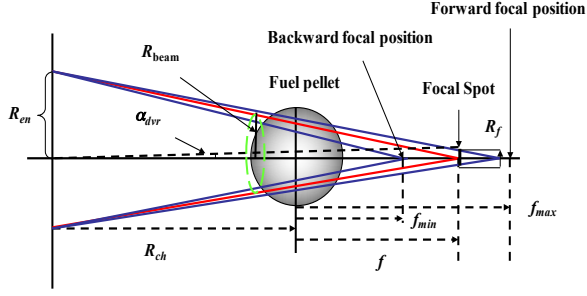

Fig. 29 The relationship between a beam emiittance and the divergence angle. Here $R_{en}$ is the HIB radius at the chamber entrance in a fusion reactor, $R_{ch}$ the reactor radius, $f$ the beam focal spot, $R_f$ the focal spot radius, $\alpha_{dvr}$ the beam divergence angle, and $R_{beam}$ is the beam radius at the target surface.

In our study, we calculate the HIB energy deposition using a stopping power. We employ the widely used expression of the HIB particle effective charge presented in Refs. 87 and 88. The stopping power of a target is the sum of the energy deposited in the target nuclei, the target bound electrons, the free electrons, and the target ions [6, 87, 88]:
$E_{stop} = E_{nuc} + E_{bound-e} + E_{free-e} + E_{ion}$:
$$E_{nucl} = 10^{-7} C_{n1}\sqrt{E_{eV}}\exp\{-45.2(C_{n2}E_{eV})^{0.277}\}$$
Here
$$C_{n1} = 4.14\times 10^6 \rho \left(\frac{A_b A_t}{A_b+A_t}\right)^{3/2}\left(\frac{Z_b Z_t}{A_t}\right)^{1/2}\left(Z_b^{3/2}+Z_t^{3/2}\right),$$
$$C_{n2} = 10^{-6}\rho\left(\frac{A_b A_t}{A_b+A_t}\right)^{3/2}(Z_b Z_t)^{-1}\left(Z_b^{3/2}+Z_t^{3/2}\right)^{-1/2}, \text{ and}$$
$E_{eV}[eV] = (eE_{beam})/A_b$, where $E_{beam}$ is the HIB ion particle energy, $A_b$ the beam ion atomic weight, $Z_b$ the beam ion atomic number, $Z_t$ the target ion atomic number and $\rho$ the target density.

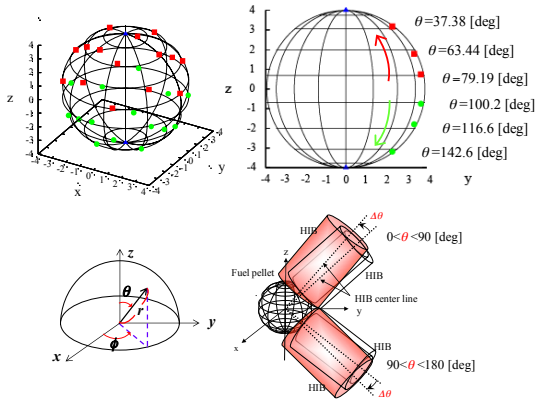

Fig. 30 Schematic diagram for the 32 HIBs illumination arrangement in Ref. [68] on the target in the case of a pellet radius of 4.0 mm. The displacement angle $\Delta\theta$ is also presented.

$E_{bound-e} = E_{Bethe} \cup E_{Lss}$. $E_{Bethe} = \frac{e^4 n_{eb}}{4\pi\varepsilon_0^2 m_e}\frac{Zeff^2}{v_{beam}^2}\left\{\log\left(\frac{2m_e v_{beam}^2}{\langle U_i\rangle}\right)\right\} - \log(1-\beta^2) - \beta^2 - \delta$ for $Z_b^{1/3} < \frac{2\varepsilon_0 hc}{e^2}\beta$, where $\delta = \sum_{n=0}^{4}a_n(\log\varepsilon_{keV})^n$, $(\log\varepsilon_{keV}) = \log(\varepsilon_{eV}\times 10^{-3})$, $\delta$ shows the shell correction, $\beta = v_{beam}/c$, $\langle U_i\rangle$ the averaged ionization potential of the target material and $v_{beam}$ the beam ion speed. The effective charge $Zeff$ of the beam ion is
$$Zeff = Z_b\left\{1 - 1.034\times\exp\left(-137.04\frac{\sqrt{\beta}}{Z_b^{0.69}c}\right)\right\}.$$
$$E_{lss} = \frac{2e^2 a_b n_{eb}}{\varepsilon_0 v_B}\frac{Z_b^{7/6}}{\left(Z_b^{3/2}+Z_t^{3/2}\right)^{3/2}}v\_beam \text{ for } Z_b^{1/3} \geq \frac{2\varepsilon_0 hc}{e^2}\beta.$$
$$E_{free-e} = \frac{e^4 n_{ef}}{4\pi\varepsilon_0^2 m_e}\frac{Zeff^2}{Vr_e^2}L_e, \text{ where}$$
$L_e = G_e(x_e)\ln(\lambda_{De}k_e) + H_e(x_e)\ln(x_e)$, $x_e = \frac{Vr_e}{V_e}$, the relative projectile speed $Vr_e = \frac{v_{beam}+\frac{4}{\pi^2}V_e}{1+\frac{4}{\pi^2}\frac{v_{beam}V_e}{c^2}}$, $V_e$ the electron thermal speed, $G_e(x) = \frac{2}{\sqrt{\pi}}\int_0^x\exp(-t^2)dt - \sqrt{\frac{2}{\pi}}x\exp\left(-\frac{x^2}{2}\right)$, and $H_e(x) = \frac{-x^3\exp(-x^2/2)}{3\sqrt{2\pi}\log x} + \frac{x^4}{x^4+12}$. The Debye shielding length is denoted by $\lambda_{De}$, $k_e = \min\left(\frac{4\pi\varepsilon_0(Vr_e^2+V_e^2)}{Zeff_e},\frac{2m_e Vr_e}{\hbar}\right)$.
$E_{ion} = \frac{e^4 n_{ei}}{4\pi\varepsilon_0^2 m_i}\frac{Zeff^2 Z_t^2}{Vr_i^2}L_i$, where $Vr_i \approx v_{beam}$, $L_i = G_i(x_i)\log(\lambda_{Di}k_i) + H_i(x_i)\log(x_i)$, $\lambda_{Di}$ the Debye length and $k_i \approx \frac{4\pi\varepsilon_0(Vr_i^2)}{Zeff}$.

We assume that the pellet is injected into a chamber vertically, and simulate the effect of a little displacement $dz$ as well as $dx$ and $dy$ on the HIB illumination non-uniformity, as shown in Fig. 31. Our illumination pattern is a basically spherically symmetric pattern.

Figure 32 shows the relation between the pellet displacement $dz$ and the non-uniformity rms for the 3.4 and 4.6 mm HIB radii in the case of the Pb+Al layer target structure. The pellet outer radius is also 4.0 mm, and the 32 HIBs are employed. For the beam radius of 4.6 mm, $\Delta\theta$ is set to be 2°. In these cases, the HIB energy deposition non-uniformity is evaluated in the Al layer in this study. The rms non-uniformity is less than 3.0% for a pellet displacement $dz$ up to about 200μm. For this pellet structure, the HIB illumination of the larger beam radius of 4.6 mm introduces a robustness against the pellet displacement $dz$. We also calculate the spectra as shown in Figs. 33 for $dz$ =100μm: (a) for the case of the beam radius of 3.4 mm at the Bragg peak layer ($R_{Bragg}$ =3.83 mm) and (b) for the beam radius of 4.6 mm and $\Delta\theta = 2°$ at the Bragg peak layer ($R_{Bragg}$ = 3.93 mm), for the global spectrum (c) for the beam radius of 3.4 mm, and (d) for the beam radius of 4.6 mm and $\Delta\theta$ is 2°. The amplitude of the spectrum mode (n, m) = (1, 0) in Fig. 33(b) becomes successfully a smaller value of $\sigma_{rms}$ = 2.73%, compared with $\sigma_{rms}$ = 4.19% shown in Fig. 33(a).

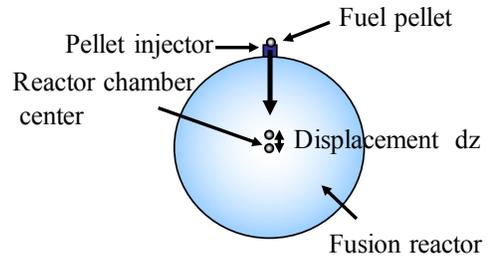

Fig. 31 Schematic diagram for the target miss alignment of $dz$.

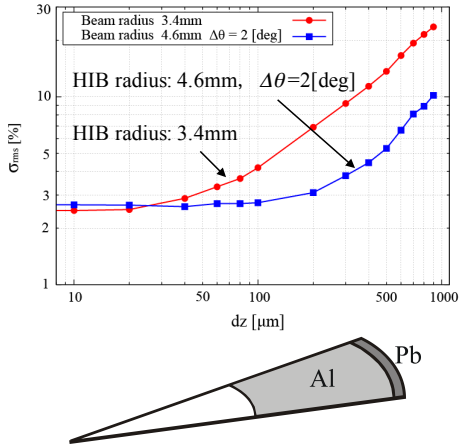

Fig. 32 The rms non-uniformity vs the pellet displacement $dz$ in a reactor chamber for the beam radiuses of 3.4 and 4.6 mm and for the pellet of the Pb+Al layer structure. The external pellet radius is 4.0 mm and the 32 HIBs are employed. When the beam radius is 4.6 mm and $\Delta\theta = 2°$, The target becomes robust against $dz$.

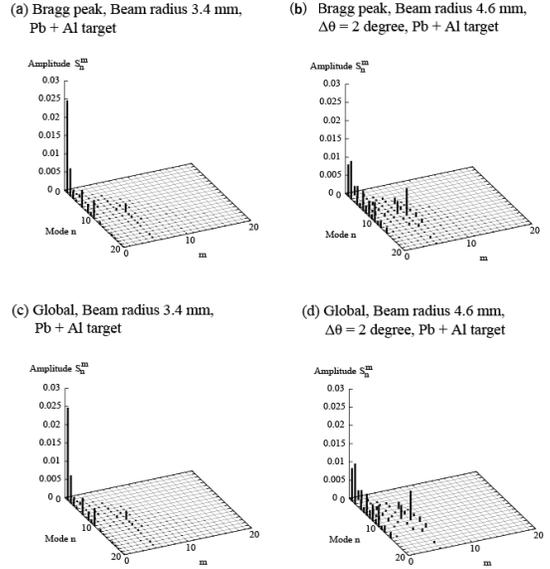

Fig. 33. The energy spectrum at the Bragg peak layer (a) for the beam radius 3.4 mm (the Bragg peak position radius $R_{Bragg}$ = 3.72 mm) and (b) for the beam radius 4.6 mm ($R_{Bragg}$ = 3.93 mm) and $\Delta\theta =2°$ for the Pb+Al layer target and the pellet dis- placement of $dz$=100 μm. The global HIB deposition energy spectrum (c) for the beam radius of 3.4 mm and (d) for the beam radius of 4.6 μm, $\Delta\theta =2°$ for the Pb+Al layer target and the pellet displacement of $dz$=100 μm.

When we perform the DT fuel target implosion fluid simulations with the target and the pulse shape in Fig. 34, the non-uniformity against $dz$ and the target gain (= Fusion output energy / HIBs input energy) are obtained as shown in Fig. 35.

When a HIB radius is larger than a target radius, outer particles of HIB do not hit the target. The HIB mishitting particles may induce another problem of an interaction between HIBs and an ICF reactor first wall, that may be a liquid metal wall. However, this interaction problem does not raise a new problem, because in a real ICF reactor the interactions of charged particles, radiation, target debris, and neutrons with the reactor first wall should be considered.

In this section, we studied the robust HIB illumination against the pellet displacement from a reactor chamber center in a direct-driven HIB ICF. In our simulation studies we found the robust HIB illumination scheme, in which the HIB radius is larger than the pellet external radius. In our study we employed the parameter of $\Delta\theta$ and optimized it to find the robust HIB illumination scheme against the pellet displacement $dz$. In our simulations the optimal $\Delta\theta$ is 2° for the beam radius of 4.6 mm and for the pellet radius of 4.0 mm. The tolerable pellet displacement is about 200–300 μm. We also analyzed the spectrum of the HIB illumination non-uniformity, and confirmed the robustness of the new illumination scheme. As described above, the HIB illumination deposition energy loss appears for a larger radius HIBs compared with the target radius. For example, the illumination energy loss is about 17.5% in the optimal case. This mishit HIB energy loss ratio may not be significant but not negligible. We also studied the HIB illumination robustness against the pellet displacements in the three-dimensional directions of *dx*, *dy*, and *dz* in Ref. [19]. The optimized 32 HIB system is also robust against the three-dimensional displacement.

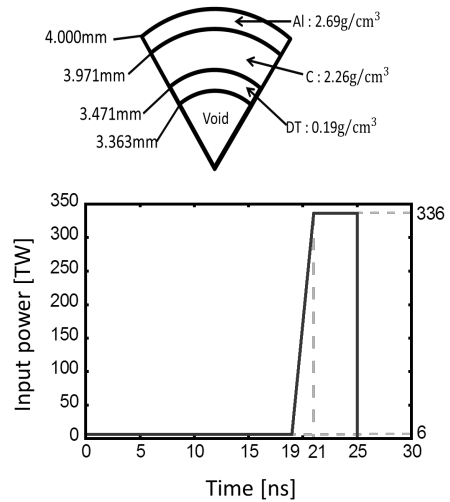

Fig. 34 An example target structure and an example HIBs pulse shape, employed to obtain the fuel target gain in Fig. 35.

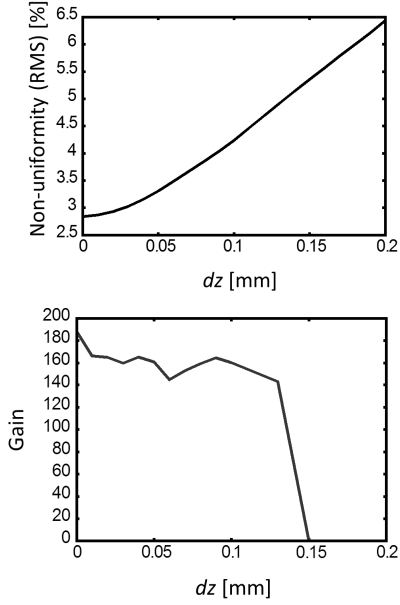

Fig. 35 The example simulation results for the HIBs illumination non-uniformity versus *dz* and for the target gain versus *dz*.

## 7 REACTOR CHABER GAS

In this section a reactor chamber gas dynamics is studied by using a simple analysis [89, 90]. A simple spherically symmetric fireball in a fusion reactor is estimated by the equation of continuity, the equation of motion and the equation of state for the adiabatic expansion:

$$\partial \rho / \partial t + (1/r^2) \partial(r^2 \rho u)/\partial r = 0,$$
$$\partial u / \partial t + u \, \partial u / \partial r = -(1/\rho) \, \partial p / \partial r,$$
$$p / \rho^\gamma = constant.$$

One of the solutions is the ZBG (Zimanyi, Bondorf and Garpman) solution [89]:

$$R^2(t) = R_0^2 + <u_t>^2 (t-t_0)^2,$$
$$\rho(t,r) = \rho_0 \left(R_0^3/R^3\right)\left(1 - r^2/R^2\right)^\alpha,$$
$$T(t,r) = T(t_0,r)\left(R_0^2/R^2\right),$$
$$\alpha = m<u_t>^2 / 2T_0 - 1.$$

In this work $\alpha=1$ is employed, and then we obtain a simple adiabatic expansion of the blast wave gas. In one shot of the DT fuel target implosion, ignition and burning, the initial temperature would be ~10keV or so. In this case the sound speed of $C_s \sim <u_t> \sim 10^8$ cm/s or so. Therefore, the blast wave traveling time $\tau$ in a fusion reactor chamber may be $\tau \sim 500 \text{cm}/10^8$ cm/s~5μsec. The target-debris chamber gas density would be the order of 1~a few torr or so, which corresponds to the gas number density of $8 \times 10^{14}/\text{cm}^3$, just after the blast wave reaches the chamber wall. Without the chamber gas supply, the expansion of the chamber gas after the target burning may make the target-debris gas dilute. During each shot, the chamber gas may be actively supplied to keep the high-density chamber gas in the chamber. When the reactor system operated with a 10~15 Hz, the blast wave propagation is too fast to give a significant interaction between the blast waves. The blast wave expands vary fast in ~5μs, and the time interval for two DT fuel target implosions is about 1/15~1/10 s.

The gas density should be high enough to compensate the focusing HIB charge (see Section 4). This is a good way for the HIB neutralization transport in a fusion reactor chamber. However, for accelerators and for the final focusing elements in the accelerator final sections, a part of the exhaust gas and debris coming up may have influences. The vacuum of the accelerator part should be kept to the low pressure to avoid the HIB ions scattering and the halo formation.

In the HIF reactor system we may have the ceramics HIB transport annular guide (see Section 4), whose inner surfaces would absorb the upcoming exhaust chamber gas to the accelerator final section. The ceramics material surface has many small holes, which absorb the debris gas and vapor [56-58, 63]. If the beam ports at the chamber wall have no mechanical shutters to stop the blast waves and the chamber gas flow, the accelerator final elements may meet the contamination. So at the final part of each accelerator near the chamber wall, each accelerator final part may several mechanical shutters, and two pairs of the mechanical shutters would confine the exhaust gas and absorb it at the ceramics guide inner surface. The absorbed gas is reused to produce a plasma at the ceramics annular guide inner surface to supply the electrons to neutralize the next HIB space charge (see Fig. 10 and its relating discussions) [56-58, 63]. This is a realistic solution to keep the accelerator upper stream parts clean from the chamber gas and debris. Figure 36 shows an example

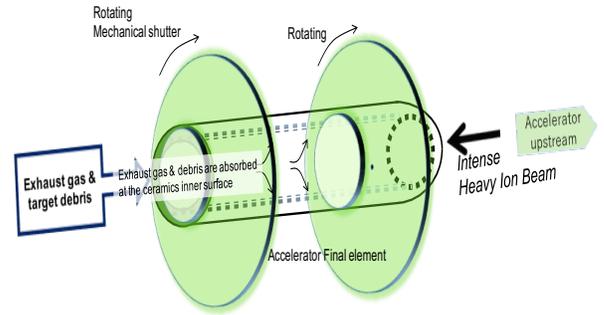

Fig. 36 At the final HIB accelerator final sections, mechanical shutters may be required to protect the accelerator upstream parts from the exhaust gas and debris from the reactor chamber. Several mechanical shutters may protect the accelerator. In this concept the ceramic annular guide in Fig. 10 has a role to absorb the exhaust gas confined between the mechanical shutters. The absorbed gas is reused to produced plasma, from which electrons are extracted to neutralize the HIB space charge just before the HIBs chamber entrance ports.

concept for the accelerator protection system.

## 8 HIF RESEARCH PROJECTS

At present an experimental device of NDCX II at Berkeley, CA, U. S. A. works on the HIB accelerator physics study [28]. In U. S. A. HIF-VNL (virtual national lab.) has been organized to conduct the HIF study experimentally and theoretically [91]. The HIF-VNL consists of Lawrence Berkeley National Laboratory (LBNL), Lawrence Livermore National Laboratory (LLNL) and Princeton university. The project overview in HIF-VNL [91] shows that they focus on the HIB accelerator physics.

The FAIR (Facility for Antiprotons and Ion Research) project has started at Darmstadt, Germany [29, 92]. FAIR is oriented to basic physics to understand the structure of matter, the evolution of the universe, etc. including the plasma physics. The HEDP (High Energy Density Physics) studies would be included in the plasma physics project in FAIR.

The HIAF (High Intensity heavy ion Accelerator Facility) in China have been planned for HIF and HEDP studies [30]. The HIAF construction site would be near Hong Kong. The total beam energy planned is ~40kJ with the pulse length of 33~130 ns. The HIAF project plan would provide a promising future accelerator for HIF studies.

In addition to the large accelerator facilities shown above, theoretical, simulation and small-scale experimental studies have been performed in various places. In Japan, historically Tokyo Institute of Technology started the HIF study in Japan 1970's, and then basic experimental studies and simulations have been intensively performed in HIF in Tokyo Institute of Technology, Utsunomiya University, Nagaoka University of Technology, Osaka University and KEK [93]. The researchers on HIF have been conducted a large number of research labs. and researchers in U. S. A., Germany, France, Russia, China, Italy, Spain, Israel, Kazakhstan, etc.

## 9 SUMMARY AND HIF ENERGY PERSPECTIVE

As we have discussed above, the ion-beam based DT inertial fusion has preferable features to release the fusion energy to sustain our human society with respect to the repetitive driver operation, the driver high efficiency, the 100% HIB energy deposition in the fuel target, the concrete HIB ion energy deposition profile in the fuel target deep layer, etc. The crucial point in ICF was the fuel implosion and ignition of the DT shell target. However, the recent NIF experiments have ensured that the DT shell target is compressed to a high density more than 1000 times of the solid density [24, 25]. The previous crucial point was solved by the NIF results. In HIF 4-5MJ of the total HIB driver energy would be needed to obtain a sufficient fusion energy output stably and repetitively.

In this sense HIF has no fatal problems to construct a HIF reactor system except technical issues relating to the T treatment, the treatment of the radioactive materials, the costs, the electric power flow of the HIF reactor plant system, etc. The previous HIF and relating reactor system designs are found in Refs. [2-4, 94]. In the near future we need to have an advanced conceptual design including the new progresses in the HIF researches as well as other remaining issues.

The intense ion beam provides also a unique tool to investigate unique physics in dynamic instability stabilization and in High Energy Density Physics, and provides a promising driver to study science in ion beam inertial fusion energy as our reliable future energy source, which is discussed and summarized in this paper. Beam ions deposit their energy inside a material, and so a warm or a high-energy area is created inside a material. One good example of the ion beam applications is an ion beam cancer therapy, in which a cancer inside a human body is directly illuminated without a serious damage of normal cells. The ion beam also has another unique feature of a precise controllability of the particle energy, the beam direction, the beam pulse duration, the beam pulse shape and also a wobbling capability by the ion beam axis precise rotation or oscillation. The wobbling capability of the ion beam presents an innovative tool to stabilize instabilities dynamically in plasmas and fluids. For example, the Rayleigh-Taylor instability growth would be controlled and stabilized significantly by the ion beam wobbling.

The precise beam control is a remarkable feature of the HIBs. The wobbling HIBs would provide a stable robust DT fuel target implosion in a fusion reactor against the non-uniformity sources as discussed above.

## Acknowledgments

The work was partly supported by JSPS, MEXT, CORE (Center for Optical Research and Education, Utsunomiya University), ASHULA, ILE / Osaka University, and CDI (Creative Department for Innovation, Utsunomiya University). The authors also would like to extend their acknowledgements to friends in HIF research group in Japan, in Tokyo Inst. of Tech., Nagaoka Univ. of Tech., KEK and also in HIF-VNL, U.S.A. and our collaborators in Utsunomiya Univ. Our former and present graduate students have contributed partly to the HIF studies, and they are also acknowledged. Especially the late Prof. Keishiro Niu, who allowed S. K. in his graduate student age to start the ion-beam-based inertial fusion instead of laser inertial fusion and magnetic confinement fusion, is acknowledged deeply. The authors would also like to express their appreciations to Prof. Yan Yun Ma in Changsha, the Editor of the Journal, who encouraged us to prepare the paper.